\documentclass[preprint,superscriptaddress,showpacs,preprintnumbers,amsmath,amssymb]{revtex4}

\usepackage{graphicx,color}
\usepackage{dcolumn}
\usepackage{bm}
\usepackage{CJK}
\usepackage{mathrsfs}

\newcommand{\lb}{\left(}
\newcommand{\rb}{\right)}
\newcommand{\ls}{\left[}
\newcommand{\rs}{\right]}

\newcommand{\ff}[1]{\frac{1}{#1}}
\newcommand{\scr}[1]{{\mathscr #1}}

\begin{document}
\begin{CJK*}{GBK}{song}

\title{Pseudospin symmetry in supersymmetric quantum mechanics:
    Schr\"odinger equations}

\author{Haozhao Liang}
 \affiliation{State Key Laboratory of Nuclear Physics and Technology, School of Physics,
Peking University, Beijing 100871, China}

\author{Shihang Shen}
 \affiliation{State Key Laboratory of Nuclear Physics and Technology, School of Physics,
Peking University, Beijing 100871, China}

\author{Pengwei Zhao}
 \affiliation{State Key Laboratory of Nuclear Physics and Technology, School of Physics,
Peking University, Beijing 100871, China}

\author{Jie Meng\footnote{Email: mengj@pku.edu.cn}}
 \affiliation{State Key Laboratory of Nuclear Physics and Technology, School of Physics,
Peking University, Beijing 100871, China}
 \affiliation{School of Physics and Nuclear Energy Engineering, Beihang University,
              Beijing 100191, China}
 \affiliation{Department of Physics, University of Stellenbosch, Stellenbosch, South Africa}

\date{\today}

\begin{abstract}
The origin of pseudospin symmetry (PSS) and its breaking mechanism are explored by combining
supersymmetry (SUSY) quantum mechanics, perturbation theory, and the similarity renormalization group (SRG) method.
The Schr\"odinger equation is taken as an example, corresponding to the lowest-order approximation in transforming a Dirac equation into a diagonal form by using the SRG.
It is shown that while the spin-symmetry-conserving term appears in the single-particle Hamiltonian $H$, the PSS-conserving term appears naturally in its SUSY partner Hamiltonian $\tilde{H}$.
The eigenstates of Hamiltonians $H$ and $\tilde{H}$ are exactly one-to-one identical except for the so-called intruder states.
In such a way, the origin of PSS deeply hidden in $H$ can be traced in its SUSY partner Hamiltonian $\tilde{H}$.
The perturbative nature of PSS in the present potential without spin-orbit term is demonstrated by the perturbation calculations, and the PSS-breaking term can be regarded as a very small perturbation on the exact PSS limits.
A general tendency that the pseudospin-orbit splittings become smaller with increasing single-particle energies can also be interpreted in an explicit way.
\end{abstract}

\maketitle

\section{Introduction}

The remarkable spin-orbit splitting for the spin doublets ($n,l,j=l\pm1/2$), i.e., the spin symmetry (SS) breaking, is one of the most important concepts for understanding the traditional magic numbers 2, 8, 20, 28, 50, and 82 for both protons and neutrons as well as 126 for neutrons in stable nuclei \cite{Haxel1949,Goeppert-Mayer1949}.
Since these magic numbers are not simply the shell closure of the harmonic oscillators, it is quite sophisticated to predict the next proton and neutron magic numbers \cite{Zhang2005,Meng2006}, which are critical for guiding the superheavy element
synthesis.
Meanwhile, the so-called pseudospin symmetry (PSS) \cite{Arima1969,Hecht1969} was introduced in 1969 to explain the near degeneracy between two single-particle states with the quantum numbers ($n-1, l + 2, j = l + 3/2$) and ($n, l, j=l + 1/2$) by defining the pseudospin doublets ($\tilde{n}=n-1,\tilde{l}=l+1,j=\tilde{l}\pm1/2$).
Based on this concept, numerous phenomena in nuclear structure have been successfully interpreted, including superdeformation \cite{Dudek1987}, identical bands \cite{Nazarewicz1990,Zeng1991}, and pseudospin partner bands \cite{Xu2008,Hua2009}, etc.
It will be quite interesting and challenging to understand the shell closure and the pseudospin symmetry on the same footing, in particular for superheavy and exotic nuclei near the limit of nucleus existence.

During the past decades, more and more exotic nuclei have became accessible with worldwide new generation radioactive ion beam (RIB) facilities.
It has been shown that the traditional magic numbers can change in the nuclei far away from the stability line~\cite{Sorlin2008}.
The splittings of both spin and pseudospin doublets play critical roles in the shell structure evolutions, for example, the $N=28$ shell closure disappears due to the quenching of the spin-orbit splitting for the $\nu1f$ spin doublets~\cite{Gaudefroy2006,Bastin2007,Tarpanov2008,Moreno-Torres2010}, the $Z=64$ sub-shell closure is closely related to
the restoration of PSS for the $\pi2\tilde{p}$ and $\pi1\tilde{f}$ pseudospin doublets~\cite{Nagai1981,Long2007,Long2009}.
Therefore, it is a fundamental task to explore the origin of SS and PSS, as well as the mechanism of their breaking in both stable and exotic nuclei.

Since the suggestion of PSS in atomic nuclei, there have been comprehensive efforts to understand its origin.
Apart from the formal relabeling of quantum numbers as shown above, various explicit transformations from the normal scheme to the pseudospin scheme have been proposed~\cite{Bohr1982,Castans1992,Blokhin1995}.
Based on the single-particle Hamiltonian of the oscillator shell model, the origin of PSS is connected with the special ratio in the
strengths of the spin-orbit and orbit-orbit interactions~\cite{Bohr1982}.
The relation between the PSS and the relativistic mean-field (RMF) theory~\cite{Serot1986,Ring1996,Vretenar2005,Meng2006} was first noted in Ref.~\cite{Bahri1992}, where the RMF theory explains approximately such special ratio in the strength of the spin-orbit and orbit-orbit interactions.

As substantial progress, the PSS was shown to be a symmetry of the Dirac Hamiltonian, where the pseudo-orbital angular momentum
$\tilde{l}$ is nothing but the orbital angular momentum of the lower component of the Dirac spinor \cite{Ginocchio1997}.
In addition, the equality in magnitude but difference in sign of the scalar potential $\mathcal{S}(\mathbf{r})$ and vector potential $\mathcal{V}(\mathbf{r})$ was suggested as the exact PSS limit by reducing the Dirac equation to a Schr\"{o}dinger-like equation \cite{Ginocchio1997}.
A more general condition $d(\mathcal{S}+\mathcal{V})/dr=0$ was proposed \cite{Meng1998}, which can be approximately satisfied in exotic nuclei with highly diffuse potentials \cite{Meng1999}.
However, since there exist no bound nuclei within such PSS limit, the non-perturbative nature of PSS in realistic nuclei has been suggested \cite{Alberto2002,Ginocchio2011}, which was also related to the consideration of the PSS as a dynamical symmetry \cite{Alberto2001}.
In this sense, an explicit and quantitative connection between the ideal PSS limits and the realistic nuclei is still missing.

After the PSS was revealed as a relativistic symmetry, numerous efforts have been dedicated to tracing the relativistic origin of PSS and its breaking mechanism in a quantitative way.
These investigations include the one-dimensional Woods-Saxon potential \cite{Panella2010},
the spherical Woods-Saxon \cite{Guo2005PLA,Xu2006,Aydogdu2010EPJA,Desplanques2010,Chen2012},
Coulomb \cite{Lisboa2003,Hamzavi2010,Castro2012},
harmonic oscillator \cite{Chen2003CPL,Lisboa2004,Ginocchio2005PRL,Guo2005NPA,Xu2007b,Marcos2008,Marcos2008b,Zhang2008},
anharmonic oscillator \cite{Zhang2009a},
Hulth\'{e}n \cite{Guo2003,Ikhdair2011AMC},
Morse \cite{Berkdemir2006,Bayrak2007,Qiang2007,Aydogdu2011,Ikhdair2011},
Rosen-Morse \cite{Oyewumi2010,Wei2010EPJA},
Eckart \cite{Jia2006,Soylu2008},
P\"{o}schl-Teller \cite{Jia2007EPJA,Jia2009,Wei2009,Candemir2012},
diatomic molecular \cite{Jia2007PS},
Manning-Rosen \cite{Wei2008,Chen2009,Wei2010PLB},
Mie-type \cite{Aydogdu2010AP,Hamzavi2010FBS},
and Yukawa \cite{Aydogdu2011PS}
potentials,
as well as the deformed
harmonic oscillator \cite{Ginocchio2004b,Guo2006PLA,Zhou2008,Zhang2009b,Zhang2009PS,Setare2010},
anharmonic oscillator \cite{Zhang2012},
Hartmann \cite{Alhaidari2006,Guo2007},
Kratzer \cite{Berkdemir2008JPA},
and Makarov \cite{Zhou2009}
potentials,
together with some formal studies \cite{Ginocchio1998PLB,Leviatan2001,Ginocchio2002,Alberto2007,Jolos2007,Jolos2012}.
Self-consistently, the PSS in spherical \cite{Gambhir1998,Ginocchio1998PRC,Marcos2000,Ginocchio2001,Marcos2001,Borycki2003,Chen2003HEPNP,Lopez-Quelle2003,Marcos2003,
Alberto2005,Marcos2005,Long2006,Xu2007a,Guo2010,Long2010}
and deformed \cite{Lalazissis1998,Sugawara-Tanabe1998,Sugawara-Tanabe2000,Sugawara-Tanabe2002,Ginocchio2004a,Sugawara-Tanabe2005,Sun2012} nuclei
have been investigated in the RMF and relativistic Hartree-Fock \cite{Bouyssy1987,Long2006RHF,Long2010RHFB} theories.
The PSS investigations were also extended to the single-particle resonances \cite{Guo2005PRC,Guo2006PRC,Zhang2007,Lu2012}
as well as the single-particle states in the Dirac sea, i.e., the SS in the anti-nucleon spectra \cite{Zhou2003,Castro2006,He2006,Liang2010,Lisboa2010}
and the $\bar\Lambda$ spectra in hyper-nuclei \cite{Song2009,Song2010,Song2011}.
The relevances of the PSS in nuclear magnetic moments and transitions \cite{Ginocchio1999PRC,Neumann-Cosel2000} as well as in nucleon-nucleus and nucleon-nucleon scatterings \cite{Ginocchio1999PRL,Leeb2000,Ginocchio2002PRCb,Leeb2004} were discussed.
The readers are referred to Refs. \cite{Ginocchio1999PR,Ginocchio2005PR} for reviews.

Recently, in Refs. \cite{Liang2011,Li2011}, the perturbation theory was used to investigate the symmetries of the Dirac Hamiltonian and their breaking in realistic nuclei.
This provides a clear way for investigating the perturbative nature of PSS.
It is found that the energy splitting of the pseudospin doublets can be regarded as a result of perturbation on the Hamiltonian with relativistic harmonic oscillator potentials, where the pseudospin doublets are degenerate.

The supersymmetric (SUSY) quantum mechanics was also used to investigate the symmetries of the Dirac Hamiltonian \cite{Leviatan2004,Typel2008,Leviatan2009,Hall2010,Zarrinkamar2010,Alhaidari2011,Zarrinkamar2011IJMPA,Zarrinkamar2011PS}.
In particular, by employing both exact and broken SUSY, the phenomenon that all states with $\tilde{l}>0$ have their own pseudospin partners except for the so-called intruder states can be interpreted naturally within a unified scheme.
A PSS breaking potential without singularity can also be obtained with the SUSY technique \cite{Typel2008}.
In contrast, singularities appear in the reduction of the Dirac equation to a Schr\"{o}dinger-like equation for the lower component of the Dirac spinor.
However, by reducing the Dirac equation to a Schr\"{o}dinger-like equation for the upper component, the effective Hamiltonian shown in Ref.~\cite{Typel2008} is not Hermitian, since the upper component wave functions alone, as the solutions of the Schr\"{o}dinger-like equation, are not orthogonal to each other.
In order to fulfill the orthonormality, an additional differential relation between the lower and upper components must be taken into account.
Thus, effectively, the upper components alone are orthogonal with respect to a modified metric.
This prevents us from being able to perform quantitative perturbation calculations.

A very recent work \cite{Guo2012} filled the gap between the perturbation calculations and the SUSY descriptions by using the similarity renormalization group (SRG) technique to transform the Dirac Hamiltonian into a diagonal form.
The effective Hamiltonian expanded in a series of $1/M$ in the Schr\"odinger-like equation thus obtained is Hermitian.
This makes the perturbation calculations possible.
Therefore, we deem it promising to understand the PSS and its breaking mechanism in a fully quantitative way by combining the SRG technique, SUSY quantum mechanics, and perturbation theory.

By using the SRG technique, a Dirac equation can be transformed into a diagonal form in a series of $1/M$, and its lowest-order approximation corresponds to a Schr\"odinger equation.
By taking this lowest-order approximation as an example, the idea for applying the SUSY quantum mechanics to trace the origin of the PSS will be illustrated and the PSS breaking mechanism will be explored quantitatively by
the perturbation theory in this paper.
In Sec.~\ref{Sect:II}, the SUSY quantum mechanics will be briefly recalled with its application for the radial Schr\"odinger equation.
The numerical details for solving the radial Schr\"odinger equation in coordinate space and the results for the single-particle eigenstates, the pseudospin-orbit splittings, the superpotentials, the PSS conserving and breaking potentials, as well as the perturbation corrections to the single-particle energies will be presented in Sec.~\ref{Sect:III}. Finally, summary and perspectives will be given in Sec.~\ref{Sect:IV}.

\section{Theoretical Framework}\label{Sect:II}

\subsection{Similarity renormalization group for the Dirac Hamiltonian}

Within the relativistic scheme, the Dirac Hamiltonian for nucleons reads
\begin{equation}
    H_D = \mathbf{\alpha}\cdot\mathbf{p} + \beta(M+\mathcal{S})+\mathcal{V},
\end{equation}
where $\mathbf{\alpha}$ and $\beta$ are the Dirac matrices, $M$ the mass of nucleon, and $\mathcal{S}$ and $\mathcal{V}$ the scalar and vector potentials, respectively.
According to the commutation and anti-commutation relations with respect to $\beta$, the Dirac Hamiltonian can be separated into the diagonal $\varepsilon$ and off-diagonal $o$ parts, $H_D=\varepsilon+o$, which satisfy
$[\varepsilon,\beta] = 0$ and $\{o,\beta\} = 0$.
In order to obtain the equivalent Schr\"odinger-like equation for nucleons, the main task is to decouple the eigenvalue equations for the upper and lower components of the Dirac spinors.
A possible way is to make the off-diagonal part of the Dirac Hamiltonian vanish with a proper unitary transformation.

According to the similarity renormalization group (SRG) technique~\cite{Wegner1994,Bylev1998}, the Hamiltonian $H_D$ is transformed by a unitary operator $U(l)$ as
\begin{equation}
    H_D(l) = U(l)H_DU^\dag(l)
\end{equation}
with $H_D(l)=\varepsilon(l)+o(l)$, $H_D(0) = H_D$, and a flow parameter $l$.
Then, the so-called Hamiltonian flow equation can be obtained by taking the differential of the above equation, i.e.,
\begin{equation}
    \frac{d}{dl}H_D(l) = [\eta(l),H_D(l)]
\end{equation}
with the anti-Hermitian generator $\eta(l) = \frac{dU(l)}{dl}U^\dag(l)$.
As pointed out in Ref.~\cite{Bylev1998}, one of the proper choices of $\eta(l)$ for letting off-diagonal part $o(l)=0$ as $l\rightarrow\infty$ reads $\eta(l) = [\beta M,H_D(l)]$.
Then, the diagonal part of the Dirac Hamiltonian $\varepsilon(l)$ at the $l\rightarrow\infty$ limit can be derived analytically in a series of $1/M$,
\begin{eqnarray}\label{Eq:Schlike}
    \varepsilon(\infty) &=& M\varepsilon_0(\infty)+\varepsilon_1(\infty)+\frac{\varepsilon_2(\infty)}{M}+\frac{\varepsilon_3(\infty)}{M^2}+\cdots\nonumber\\
    &=&
    \beta M+(\beta\mathcal{S}+\mathcal{V})+\ff{2M}\beta (\mathbf{\alpha}\cdot\mathbf{p})^2\nonumber\\
    &&+\ff{8M^2} \ls\ls \mathbf{\alpha}\cdot\mathbf{p},(\beta\mathcal{S}+\mathcal{V})\rs, \mathbf{\alpha}\cdot\mathbf{p}\rs+\cdots
\end{eqnarray}
In such way, the eigenequations for the upper and lower components of the Dirac spinors are completely decoupled.
The equivalent Sch\"ordinger-like equations for nucleons with Hermitian effective Hamiltonians can be obtained.
The corresponding details can be found in Ref. \cite{Guo2012}.

For the nucleons in the Fermi sea, the eigenequations with the Hamiltonian in Eq. (\ref{Eq:Schlike}) up to the $1/M$ order correspond to Sch\"ordinger equations
\begin{equation}
    \ls -\ff{2M}\nabla^2+V(\mathbf{r})\rs\psi_\alpha(\mathbf{r}) = E_\alpha\psi_\alpha(\mathbf{r}),
\end{equation}
where $V(\mathbf{r})=\mathcal{S}(\mathbf{r})+\mathcal{V}(\mathbf{r})$, and the rest mass of nucleon $M$ is reduced in the single-particle energies $E$.
By assuming the spherical symmetry, the radial equations can be cast in the form
\begin{equation}\label{Eq:Schr}
    H R_a(r) = E_a R_a(r)
\end{equation}
with the single-particle Hamiltonian
\begin{equation}\label{Eq:Hka}
    H = -\frac{d^2}{2Mdr^2} + \frac{\kappa_a(\kappa_a+1)}{2Mr^2} + V(r),
\end{equation}
and the single-particle wave functions
\begin{equation}
    \psi_\alpha(\mathbf{r}) = \frac{R_a(r)}{r}\scr Y_{j_a m_a}^{l_a}(\hat{\mathbf{r}}),
\end{equation}
where $\scr Y^{l_a}_{j_am_a}$ are the spherical harmonics spinors, the single-particle eigenstates are specified by the set of quantum numbers $\alpha=(a, m_a)=(n_a, l_a, j_a, m_a)$,
and the good quantum number $\kappa=\mp(j+1/2)$ for $j=l\pm1/2$  is adopted.

It can be clearly seen that $H$ conserves the explicit spin symmetry (SS) for the spin doublets $a$ and $b$ with $\kappa_a+\kappa_b=-1$, which leads to the same centrifugal barrier (CB) $\kappa(\kappa+1)/(2Mr^2)$.
Similarly, in order to investigate the origin of the pseudospin symmetry (PSS) and its breaking mechanism, it is crucial to identify the corresponding term proportional to $\tilde{l}(\tilde{l}+1)=\kappa(\kappa-1)$, which leads to the same $\kappa(\kappa-1)$ values for the pseudospin doublets $a$ and $b$ with $\kappa_a+\kappa_b=1$.
As noted in Ref.~\cite{Typel2008}, we also consider the supersymmetric (SUSY) quantum mechanics is one of the most promising approaches for identifying such $\kappa(\kappa-1)$ structure.

In the following, we will briefly recall the key formalism of SUSY quantum mechanics \cite{Cooper1995,Typel2008}.
Then we will focus on the application of the SUSY quantum mechanics to the Schr\"{o}dinger equations.

\subsection{Supersymmetric quantum mechanics}

It has been shown that every second-order Hamiltonian can be factorized in a product of two Hermitian conjugate first-order operators~\cite{Infeld1951}, i.e.,
\begin{equation}
    H_1=B^+B^-
\end{equation}
with $B^-=[B^+]^\dag$.
Its SUSY partner Hamiltonian can thus be constructed by~\cite{Cooper1995}
\begin{equation}
    H_2=B^-B^+.
\end{equation}
Since the extended SUSY Hamiltonian
\begin{equation}
    H_S=
    \lb\begin{array}{cc}
        H_1 & 0 \\ 0 & H_2
    \end{array}\rb
\end{equation}
is the square of the Hermitian operators
\begin{equation}
    H_S=
    \lb\begin{array}{cc}
        0 & B^+ \\ B^- & 0
    \end{array}\rb^2
    =
    \lb\begin{array}{cc}
        0 & -iB^+ \\ iB^- & 0
    \end{array}\rb^2,
\end{equation}
all eigenvalues $E_S(n)$ of the eigenvalue equation
\begin{equation}\label{Eq:SUSYeq}
    H_S\Psi_S(n)=E_S(n)\Psi_S(n)
\end{equation}
are non-negative, and the two-component wave functions read
\begin{equation}
    \Psi_S(n)=
    \lb\begin{array}{c}
        \psi_1(n) \\ \psi_2(n)
    \end{array}\rb,
\end{equation}
where $\psi_1(n)$ and $\psi_2(n)$ are the eigenfunctions of $H_1$ and $H_2$, respectively.
It can be easily seen that, for each eigenstate with $E_S(n)>0$, it is an eigenstate for both $H_1$ and $H_2$, and the corresponding eigenfunctions satisfy
\begin{equation}
    \psi_2(n)=\frac{B^-}{\sqrt{E_S(n)}}\psi_1(n),\quad
    \psi_1(n)=\frac{B^+}{\sqrt{E_S(n)}}\psi_2(n)
\end{equation}
with the normalization factors $1/\sqrt{E_S(n)}$.

\begin{figure}
\includegraphics[width=8cm]{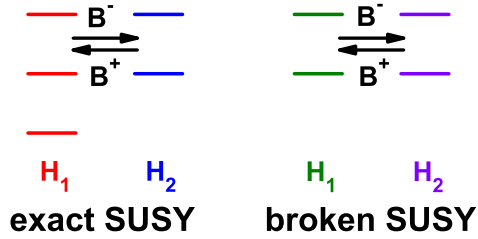}
\caption{(Color online) Schematic patterns of the exact and broken supersymmetries (SUSY).
    \label{Fig1}}
\end{figure}

The property of SUSY can be either exact (also called unbroken) or broken~\cite{Cooper1995}.
On one hand, the SUSY is exact when the eigenvalue equation (\ref{Eq:SUSYeq}) has a zero energy eigenstate $E_S(0)=0$. In this case, as an usual convention, the Hamiltonian $H_1$ has an additional eigenstate at zero energy that does not appear in its partner Hamiltonian $H_2$, since
\begin{equation}
    B^-\psi_1(0)=0, \quad \psi_2(0)=0,
\end{equation}
i.e., the trivial eigenfunction of $H_2$ identically equals to zero.
It is noted that, for the systems with periodic potentials, the exact SUSY can have a pair of ground-states with zero energy \cite{Braden1985,Dunne1998,Correa2008PRL,Correa2008JPA}.
On the other hand, the SUSY is broken when the eigenvalue equation (\ref{Eq:SUSYeq}) does not have any zero energy eigenstate. In this case, the partner Hamiltonians $H_1$ and $H_2$ have identical spectra.
The schematic patterns of both cases are illustrated in Fig.~\ref{Fig1}.

\subsection{SUSY quantum mechanics for Schr\"{o}dinger equations}\label{Sect:SUSYfor}

For applying the SUSY quantum mechanics to the Schr\"{o}dinger equations shown in Eq.~(\ref{Eq:Schr}), first of all, one sets a couple of Hermitian conjugate first-order operators as
\begin{equation}
    B^+_\kappa = \ls Q_\kappa(r)-\frac{d}{dr}\rs\ff{\sqrt{2M}},
    \quad
    B^-_\kappa = \ff{\sqrt{2M}}\ls Q_\kappa(r)+\frac{d}{dr}\rs,
\end{equation}
where the $Q_\kappa$ as the function of $r$ are the so-called superpotentials to be determined~\cite{Cooper1995,Typel2008}.
Then, the SUSY partner Hamiltonians $H_1$ and $H_2$ can be constructed as
\begin{subequations}
\begin{eqnarray}
    H_1(\kappa) &=& B^+_\kappa B^-_\kappa = \ff{2M}\ls-\frac{d^2}{dr^2}+Q_\kappa^2-Q'_\kappa\rs,\\
    H_2(\kappa) &=& B^-_\kappa B^+_\kappa = \ff{2M}\ls-\frac{d^2}{dr^2}+Q_\kappa^2+Q'_\kappa\rs.
\end{eqnarray}
\end{subequations}
In order to explicitly identify the $\kappa(\kappa+1)$ structure shown in Eq.~(\ref{Eq:Hka}), the reduced superpotentials $q_\kappa(r)$ are assumed as \cite{Typel2008}
\begin{equation}
    q_\kappa(r) = Q_\kappa(r) - \frac{\kappa}{r}.
\end{equation}
In such way, the Hamiltonians $H_1$ and $H_2$ can be further expressed as
\begin{widetext}
\begin{subequations}\label{Eq:BB}
\begin{eqnarray}
    H_1(\kappa) &=& B^+_\kappa B^-_\kappa = \ff{2M}\ls-\frac{d^2}{dr^2}+\frac{\kappa(\kappa+1)}{r^2}+q_\kappa^2+\frac{2\kappa}{r}q_\kappa-q'_\kappa\rs,\label{Eq:BB1}\\
    H_2(\kappa) &=& B^-_\kappa B^+_\kappa = \ff{2M}\ls-\frac{d^2}{dr^2}+\frac{\kappa(\kappa-1)}{r^2}+q_\kappa^2+\frac{2\kappa}{r}q_\kappa+q'_\kappa\rs.\label{Eq:BB2}
\end{eqnarray}
\end{subequations}
\end{widetext}
It can be seen that not only does the $\kappa(\kappa+1)$ structure appear in $H_1$ but also the $\kappa(\kappa-1)$ structure explicitly appears in the SUSY partner Hamiltonian $H_2$.
The so-called pseudo-centrifugal barrier (PCB) term $\kappa(\kappa-1)/(2Mr^2)$ leads to the conservation of the PSS.

In general, the Hamiltonian $H$ in the Schr\"odinger equation (\ref{Eq:Schr}) differs from the SUSY Hamiltonian $H_1$ in Eq.~(\ref{Eq:BB}) by a constant, i.e.,
\begin{equation}\label{Eq:Eshift}
    H(\kappa) = H_1(\kappa)+e(\kappa),
\end{equation}
where $e(\kappa)$ is the so-called energy shift~\cite{Cooper1995,Typel2008}.
The $\kappa$-dependent energy shifts can be determined in the following ways:
1) For the case of $\kappa<0$, it is known that the most deeply bound state for a given $\kappa$, e.g, $1s_{1/2}$, $1p_{3/2}$, etc., has no pseudospin partner. This indicates the exact SUSY is achieved, and requires
\begin{equation}
    e(\kappa) = E_{1\kappa}.
\end{equation}
2) For the case of $\kappa>0$, each single-particle state has its own pseudospin partner. This indicates the SUSY is broken, and thus the corresponding energy shift can be, in principle, any number which makes the whole set of $H_1$ eigenstates positive.
In practice, the energy shifts in this case are determined by assuming that the pseudospin-orbit (PSO) potentials vanish as $r\rightarrow0$.
This vanishing behavior is similar to that of the usual surface-peak-type spin-orbit (SO) potentials.

In order to fulfill the above condition, we first analyze the asymptotic behaviors of the reduced superpotentials $q_\kappa(r)$.
Combining Eqs.~(\ref{Eq:Hka}), (\ref{Eq:BB1}), and (\ref{Eq:Eshift}), one has
\begin{equation}\label{Eq:HV}
    \ff{2M}\ls q_\kappa^2(r)+\frac{2\kappa}{r}q_\kappa(r)-q'_\kappa(r)\rs + e(\kappa) = V(r).
\end{equation}
At large radius, for potential $\lim_{r\rightarrow\infty}V(r)=0$, $q_\kappa(r)$ becomes a constant as
\begin{equation}\label{Eq:qinf}
    \lim_{r\rightarrow\infty}q_\kappa(r) = \sqrt{-2Me(\kappa)}.
\end{equation}
At small radius, for any regular potential $V(r)$, it requires $q_\kappa(0)=0$, and also
\begin{equation}\label{Eq:q0}
    \lim_{r\rightarrow0}q_{\kappa}(r)=\frac{2M(e(\kappa)-V)}{(1-2\kappa)}r
\end{equation}
as a linear function of $r$.

As the PSO potentials vanish at the original point, it requires that $\lim_{r\rightarrow0}q_{\kappa_a}(r) = \lim_{r\rightarrow0}q_{\kappa_b}(r)$ with $\kappa_a+\kappa_b=1$ for pseudospin doublets~\cite{Typel2008}.
Finally, the energy shifts are determined by
\begin{equation}
    e(\kappa_a) = 2\left.V\right|_{r=0} - e(\kappa_b)
\end{equation}
for the case of $\kappa_a>0$.

Before ending this section, it is interesting to seek a possible exact PSS limit analytically.
First of all, combining Eqs.~(\ref{Eq:Hka}), (\ref{Eq:BB2}), and (\ref{Eq:Eshift}), the SUSY partner Hamiltonian of $H(\kappa)$ reads
\begin{equation}\label{Eq:Htil}
    \tilde{H}(\kappa)=H_2(\kappa) + e(\kappa)
    = -\frac{d^2}{2Mdr^2} + \frac{\kappa(\kappa-1)}{2Mr^2} + \tilde{V}_\kappa(r)
\end{equation}
with
\begin{equation}\label{Eq:Vtil}
    \tilde{V}_\kappa(r)=V(r)+q'_\kappa(r)/M.
\end{equation}
In this paper, we use a tilde to denote the operators, potentials, and wave functions belonging to the representation of $\tilde{H}$.
Then, by definition, the exact PSS limit holds $E_{n\kappa_a} = E_{(n-1) \kappa_b}$ with $\kappa_a<0$ and $\kappa_a+\kappa_b=1$. This indicates
\begin{equation}
    H_2(\kappa_a) + e(\kappa_a) = H_2(\kappa_b) + e(\kappa_b).
\end{equation}
By combining Eqs.~(\ref{Eq:BB}) and (\ref{Eq:HV}), as well as the boundary condition $q_\kappa(0)=0$, one can readily have
\begin{equation}
    q_{\kappa_a}(r) = q_{\kappa_b}(r) = M\omega_{\kappa} r
\end{equation}
with a known constant $\omega_{\kappa} \equiv (e(\kappa_a)-e(\kappa_b))/(\kappa_b-\kappa_a)$.
As the reduced superpotentials $q_\kappa(r)$ are linear functions of $r$, the central potential $V(r)$ in $H$ has the form
\begin{equation}
    V_{\rm HO}(r) = \frac{M}{2}\omega_{\kappa}^2 r^2+V(0).
\end{equation}
The corresponding PSS limit is nothing but the well known case with harmonic oscillator (HO) potentials,
which leads to the energy degeneracy of the whole major shell.

\section{Results and discussion}\label{Sect:III}

\begin{figure}
\includegraphics[width=8cm]{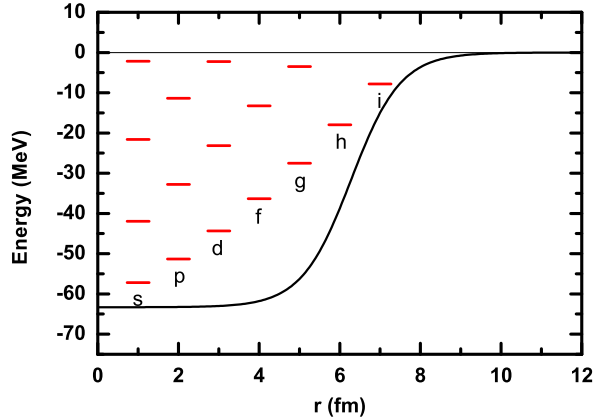}
\caption{(Color online) Woods-Saxon potential and discrete eigenstates for neutrons in the nucleus $^{132}$Sn.
    \label{Fig2}}
\end{figure}

To perform a quantitative investigation, the mass of nucleon is taken as $M=939.0$~MeV, and the central potential $V(r)$ in Eq.~(\ref{Eq:Hka}) is chosen as a Woods-Saxon form
\begin{equation}
    V(r) = \frac{V_0}{1+e^{(r-R)/a}}
\end{equation}
with the parameters $V_0=-63.297$~MeV, $R=6.278$~fm, and $a=0.615$~fm, which corresponds to the neutron potential provided in Ref.~\cite{Koepf1991} by taking $N=82$ and $Z=50$.
This potential is shown as the solid line in Fig.~\ref{Fig2}.
The radial Schr\"odinger equations are solved in coordinate space by the shooting method~\cite{Meng1998Nucl.Phys.A} within a spherical box with radius $R_{\rm box} = 20$~fm and mesh size $dr = 0.05$~fm.

\subsection{Representation of single-particle Hamiltonian $H$}

\begin{figure}
\includegraphics[width=8cm]{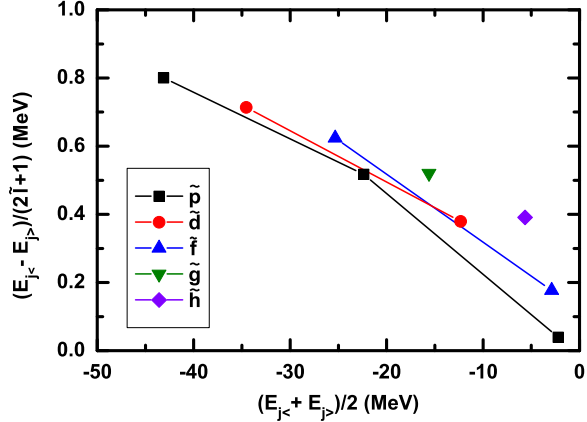}
\caption{(Color online) Pseudospin-orbit splittings $(E_{j_<}-E_{j_>})/(2\tilde{l}+1)$ versus the average single-particle energies $(E_{j_<}+E_{j_>})/2$ for pseudospin doublets.
    \label{Fig3}}
\end{figure}

In Fig.~\ref{Fig2}, the discrete single-particle states obtained in the Woods-Saxon potential are shown.
In order to see the $\kappa$-dependence and the energy-dependence of pseudospin-orbital (PSO) splittings more clearly, we plot the reduced PSO splittings $(E_{j_<}-E_{j_>})/(2\tilde{l}+1)$ versus their average  single-particle energies $E_{\rm av}=(E_{j_<}+E_{j_>})/2$ in Fig.~\ref{Fig3}, where $j_<$ ($j_>$) denotes the states with $j=\tilde{l}-1/2$ ($j=\tilde{l}+1/2$).
It is seen that the amplitudes of the reduced PSO splittings are less than 1~MeV.
Moreover, as a general tendency, the splittings become smaller with the increasing single-particle energies.
Such energy-dependent behavior was also found in the self-consistent relativistic continuum Hartree-Bogoliubov (RCHB) calculations~\cite{Meng1998,Meng1999}.
It was also reported that the PSO splittings can even reverse in resonance states~\cite{Guo2005PRC,Guo2006PRC,Zhang2007}.
Note that the spin-orbit (SO) splittings never reverse in realistic nuclei.
Therefore, it is very interesting to investigate the physical mechanism for such energy-dependent behavior.
This will also help us to figure out whether or not the pseudospin symmetry (PSS) is an accidental symmetry \cite{Marcos2008}.

\begin{figure}
\includegraphics[width=8cm]{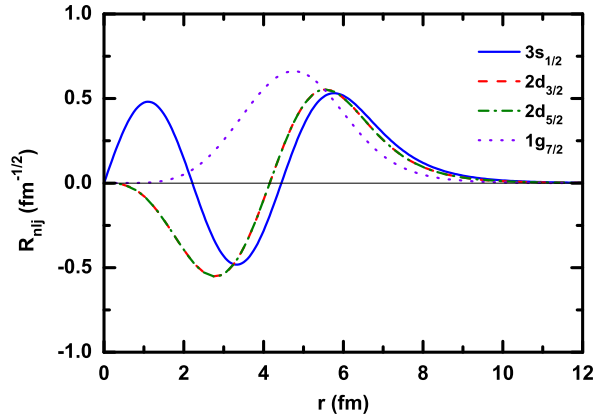}
\caption{(Color online) Single-particle wave functions $R_{nlj}(r)$ of $H$ for the $3s_{1/2}$, $2d_{3/2}$, $2d_{5/2}$, and $1g_{7/2}$ states.
    \label{Fig4}}
\end{figure}

In Fig.~\ref{Fig4}, the corresponding single-particle radial wave functions $R_{nlj}(r)$ of $H$ are shown by taking the $3s_{1/2}$, $2d_{3/2}$, $2d_{5/2}$, and $1g_{7/2}$ states, i.e., the pseudospin doublets $2\tilde{p}$ and $1\tilde{f}$, as examples.
In the following discussions, these two pairs of pseudospin doublets will be often used for illustration.
It is clear that the wave functions of the spin doublets are identical since there is no spin-orbit term in $H$.
In contrast, the wave functions of the pseudospin doublets are very different from each other.
This leads to difficulty in analyzing the origin of the PSS and its breaking.

\begin{table}
\caption{Contributions from the kinetic term (kin.), centrifugal barrier (CB), and central potential (cen.) to the single-particle energies $E$ and the corresponding pseudospin-orbit splittings $\Delta E_{\rm PSO}$ for the pseudospin doublets $2\tilde{p}$ and $1\tilde{f}$.
All units are in MeV.}
\label{Tab1}
\begin{ruledtabular}
    \begin{tabular}{lrrrr}
            state & \multicolumn{1}{c}{$E_{\rm kin.}$} & \multicolumn{1}{c}{$E_{\rm CB}$} & \multicolumn{1}{c}{$E_{\rm cen.}$} & \multicolumn{1}{c}{$E$} \\ \hline
            $3s_{1/2}$ & 28.953 &   0.000 & -50.545 & -21.591 \\
            $2d_{3/2}$ & 16.845 &  11.758 & -51.746 & -23.143 \\
  $\Delta E_{\rm PSO}$ & 12.109 & -11.758 &   1.201 &   1.552 \\ \hline
            $2d_{5/2}$ & 16.845 &  11.758 & -51.746 & -23.143 \\
            $1g_{7/2}$ &  6.197 &  20.483 & -54.188 & -27.508 \\
  $\Delta E_{\rm PSO}$ & 10.648 &  -8.725 &   2.442 &   4.365 \\
    \end{tabular}
\end{ruledtabular}
\end{table}

Prior to the quantitative analysis by using the perturbation theory in Ref.~\cite{Liang2011}, the investigation of PSO splittings $\Delta E_{\rm PSO}$ was usually done by decomposing the contributions by terms, where each contribution is calculated with the corresponding operator $\hat O_i$ by
\begin{equation}
    E_i = \int R(r)\hat O_i R(r) dr.
\end{equation}

Within the representation of $H$ shown in Eq.~(\ref{Eq:Hka}), the operators of the kinetic term, centrifugal barrier (CB), and central potential read $-d^2/(2Mdr^2)$, $\kappa(\kappa+1)/(2Mr^2)$, and $V(r)$, respectively.
In Table~\ref{Tab1}, the contributions from these terms to the single-particle energies $E$ as well as the corresponding PSO splittings $\Delta E_{\rm PSO}$ are shown for the pseudospin doublets $2\tilde{p}$ and $1\tilde{f}$.
It is not surprising that, within this representation, the contributions to $\Delta E_{\rm PSO}$ come from all channels, while they substantially cancel to each other in a sophisticated way.

In previous studies, the phenomenon of such strong cancellations among different terms was usually associated with the dynamical~\cite{Alberto2001,Marcos2008,Guo2012} and even the nonperturbative~\cite{Alberto2002,Lisboa2010,Ginocchio2011} nature of PSS.
However, such connection is mystified and sometimes even misleading.
We will demonstrate by using the perturbation theory that the nature of PSS in the present investigation is indeed perturbative.

The main idea for using the Rayleigh-Schr\"odinger perturbation theory~\cite{Greiner1994} to investigate the spin symmetry (SS) and PSS in single-particle Hamiltonian as well as their breaking in atomic nuclei can be found in Ref.~\cite{Liang2011}.
Following this idea, the Hamiltonian $H$ is split as
\begin{equation}
    H = H_0 + W,
\end{equation}
where $H_0$ conserves the exact PSS and $W$ is identified as the corresponding symmetry breaking
potential. The condition
\begin{equation}\label{Eq:criterion}
    \left|\frac{W_{mk}}{E_k-E_m}\right|\ll 1
    \quad\mbox{for}\quad m\neq k
\end{equation}
with $W_{mk}=\left< \psi_m \right| W \left| \psi_k \right>$ determines whether $W$ can be treated as a small perturbation and governs the convergence of the perturbation series~\cite{Greiner1994}.

\begin{figure}
\includegraphics[width=8cm]{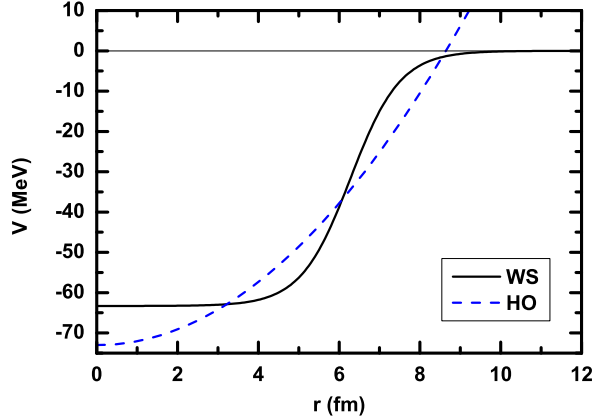}
\caption{(Color online) Woods-Saxon potential in $H$ (solid line) and harmonic oscillator potential in $H^{\rm HO}_0$ (dashed line)
as a function of $r$.
    \label{Fig5}}
\end{figure}

For the present case, it has been analytically shown in Section~\ref{Sect:SUSYfor} that the Hamiltonian with harmonic oscillator (HO) potentials is one of the exact PSS limits.
Thus, one has
\begin{equation}
    H^{\rm HO}_0 = -\ff{2M}\ls\frac{d^2}{dr^2} + \frac{\kappa(\kappa+1)}{r^2}\rs + \frac{M}{2}\omega^2 r^2+V(0),
\end{equation}
and $W^{\rm HO}$ is just the difference between $H$ and $H^{\rm HO}_0$.
To minimize the perturbations to the $sdg$ states, the coefficient $\omega$ is chosen as $1.118\times41 A^{-1/3}$~MeV, and the trivial constant $V(0)$ is taken as $-73$~MeV, as illustrated in Fig.~\ref{Fig5}.
Although the symmetry breaking potential $W^{\rm HO}$ diverges at $r\rightarrow\infty$ due to the parabolic behavior of $H^{\rm HO}_0$, the property that the bound state wave functions decay exponentially at large radius leads to convergent results of the matrix elements $W_{mk}$.

\begin{figure}
\includegraphics[width=8cm]{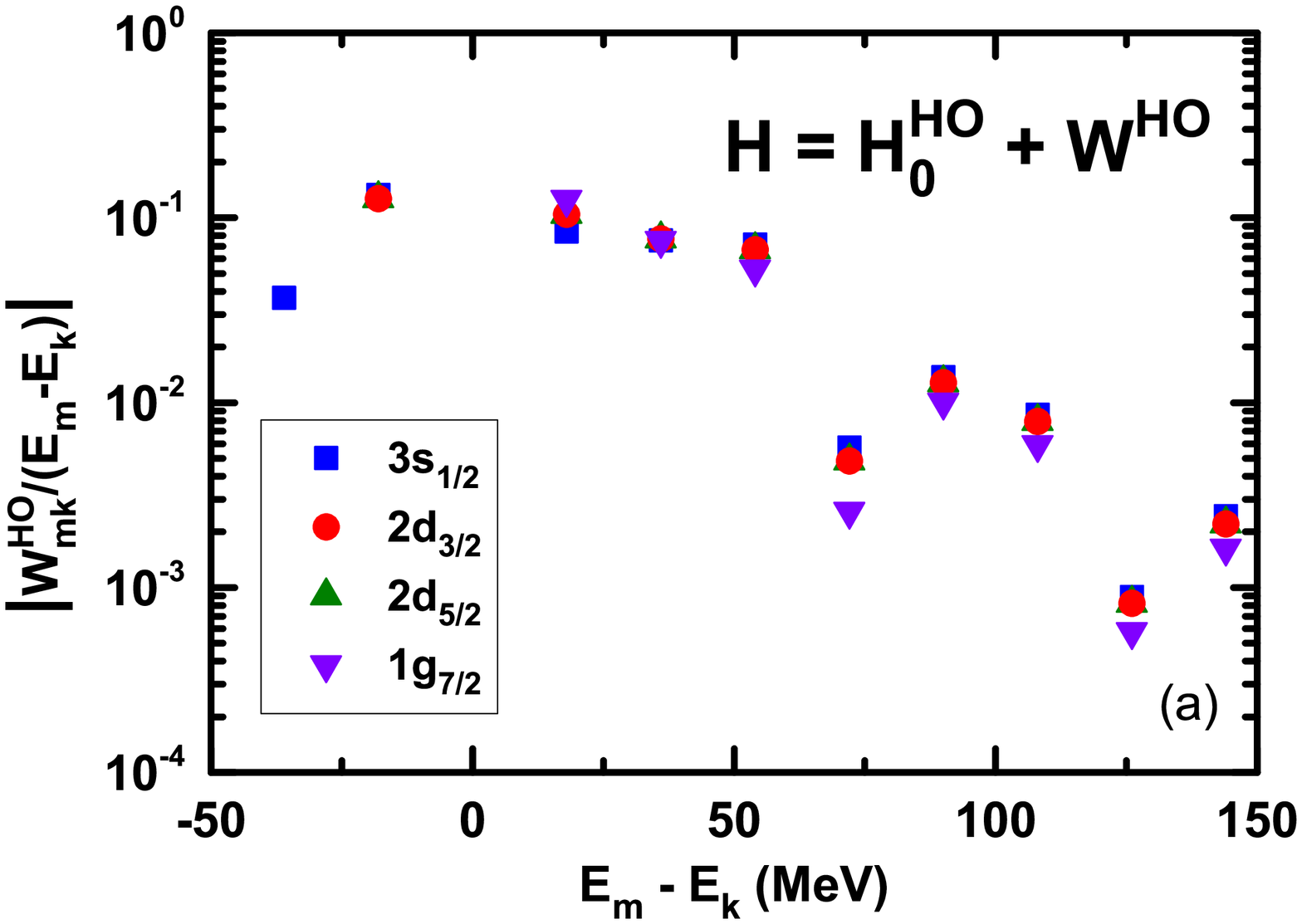}\\
\includegraphics[width=7.55cm]{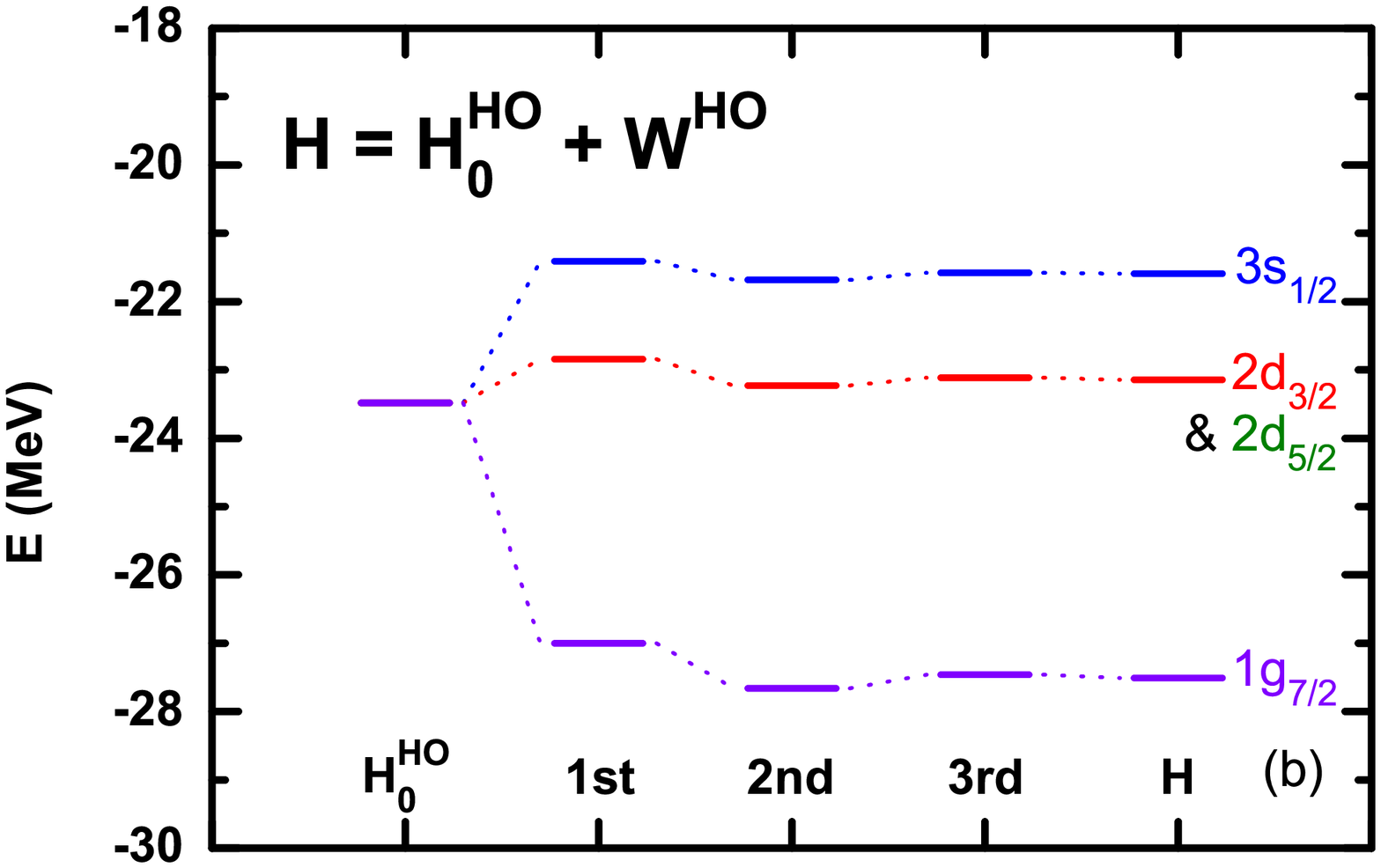}
\caption{(Color online)
Panel (a): Values of $|W^{\rm HO}_{mk}/(E_m-E_k)|$ versus the energy differences $E_m-E_k$ for the pseudospin doublets $k = 2\tilde{p}$ and $1\tilde{f}$.
Panel (b): Corresponding single-particle energies obtained at the exact PSS limit $H^{\rm HO}_0$, and by the first-, second-, and third-order perturbation calculations, as well as those obtained with $H$.
    \label{Fig6}}
\end{figure}

In the upper panel of Fig.~\ref{Fig6}, the values of $\left|W^{\rm HO}_{mk}/(E_m-E_k)\right|$ for the pseudospin doublets $k = 2\tilde{p}$ and $1\tilde{f}$ are shown as functions of the energy differences $E_m-E_k$, where the unperturbed eigenstates in perturbation calculations are chosen as those of $H_0^{\rm HO}$.
Since the spherical symmetry is assumed, only the single-particle states $m$ and $k$ with the same quantum number $\kappa$ lead to non-vanishing matrix elements $W_{mk}$. Similar to the results in Ref.~\cite{Liang2011}, the values of $\left|W^{\rm HO}_{mk}/(E_m-E_k)\right|$ decrease as a general tendency when the energy differences $E_m-E_k$ increase.
From the mathematical point of view, this property provides natural cut-offs of the single-particle states in perturbation calculations.
For example, the first-, second-, and third-order perturbation corrections to the single-particle energies in the present calculations are of $0.001$~MeV accuracy when the energy cut-off for $E_m$ is taken as $150$~MeV.
Furthermore, it is shown that the largest perturbation correction is roughly 0.13.
This indicates that the criterion in Eq.~(\ref{Eq:criterion}) can be fulfilled.

In the lower panel of Fig.~\ref{Fig6}, the single-particle energies obtained at the exact PSS limit $H^{\rm HO}_0$, and their counterparts obtained by the first-, second-, and third-order perturbation calculations, as well as those obtained with $H$ are shown from left to right.
Although the perturbation corrections do not converge very fast since the largest perturbations for all four states are beyond 0.1, the PSO splittings are well reproduced up to the third-order perturbation calculations.

The above results have pinned down the perturbative nature of PSS in the present investigation.
In the next subsection, we will study the origin of the PSS and its breaking mechanism in an explicit way within the representation of the supersymmetric (SUSY) partner Hamiltonian $\tilde{H}$.

\subsection{Representation of the SUSY partner Hamiltonian $\tilde{H}$}

\begin{figure}
\includegraphics[width=8cm]{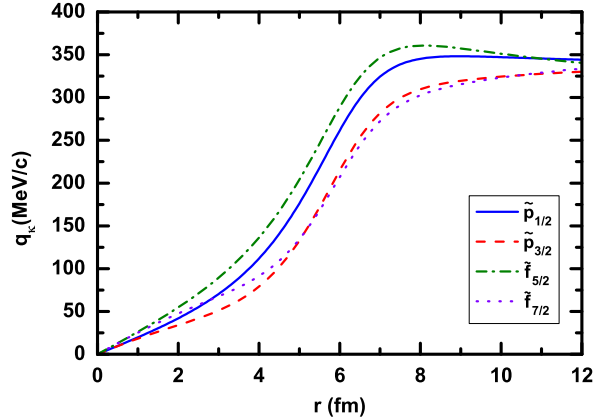}
\caption{(Color online)
Reduced superpotentials $q_\kappa(r)$ for the $\tilde{p}$ and $\tilde{f}$ blocks.
    \label{Fig7}}
\end{figure}

In order to obtain the SUSY partner Hamiltonian $\tilde{H}$ in Eq.~(\ref{Eq:Htil}), first of all, one should solve the first-order differential equation (\ref{Eq:HV}) for the reduced superpotentials $q_\kappa(r)$ with the boundary condition $q_\kappa(0)=0$.
By taking the $\tilde{p}$ and $\tilde{f}$ blocks as examples, the corresponding $q_\kappa(r)$ are shown in units of MeV/$c$ in Fig.~\ref{Fig7}.
Since the L.H.S. of Eq.~(\ref{Eq:HV}) contains a $\kappa$-dependent term $\propto\kappa/r$, the reduced superpotentials $q_\kappa(r)$ thus obtained are also $\kappa$-dependent.
In contrast, it should be emphasized that $q_\kappa(r)$ does not depend on the radial quantum number $n$ for a given $\kappa$.
One will discover in the following that such an $n$-independent property is essential for understanding the general pattern of $\Delta E_{\rm PSO}$ versus $E_{\rm av}$ as shown in Fig.~\ref{Fig3}.
In addition, it can also be examined that the reduced superpotentials $q_\kappa(r)$ satisfy their asymptotic behaviors at $r\rightarrow0$ and $r\rightarrow\infty$ in Eqs.~(\ref{Eq:q0}) and (\ref{Eq:qinf}).

The $\kappa$-dependent central potentials $\tilde V_\kappa(r)$ in $\tilde{H}$ can be then calculated by combining Eqs.~(\ref{Eq:HV}) and (\ref{Eq:Vtil}). Then, the corresponding asymptotic behaviors read
\begin{equation}
    \lim_{r\rightarrow0}\tilde{V}_{\kappa}(r)=V+\frac{2(e(\kappa)-V)}{(1-2\kappa)}
\end{equation}
and
\begin{equation}
    \lim_{r\rightarrow\infty}\tilde{V}_{\kappa}(r) = 0.
\end{equation}
It is important that these potentials are regular and converge at both $r\rightarrow0$ and $r\rightarrow\infty$.

\begin{figure*}
\includegraphics[width=8cm]{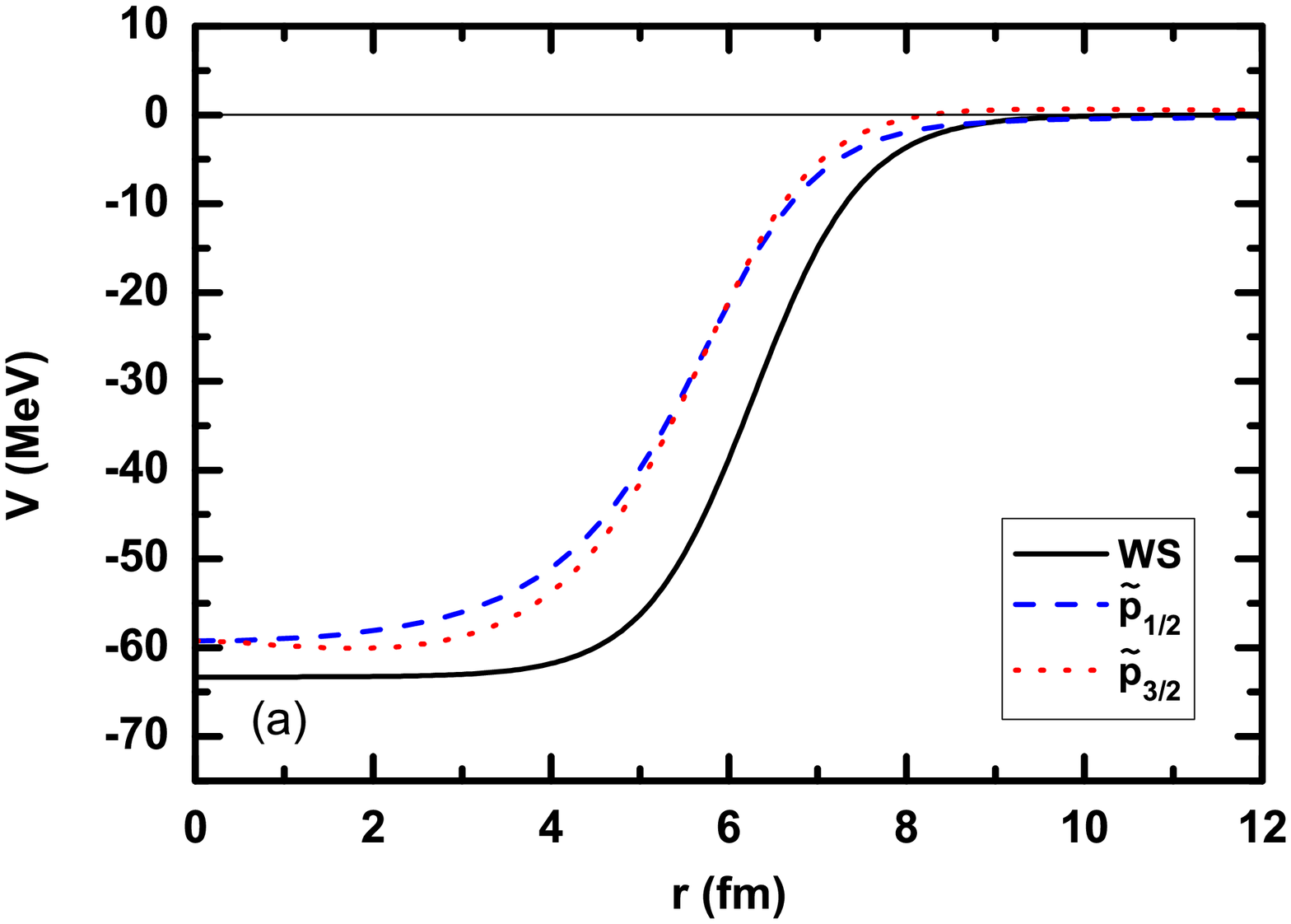}
\includegraphics[width=8cm]{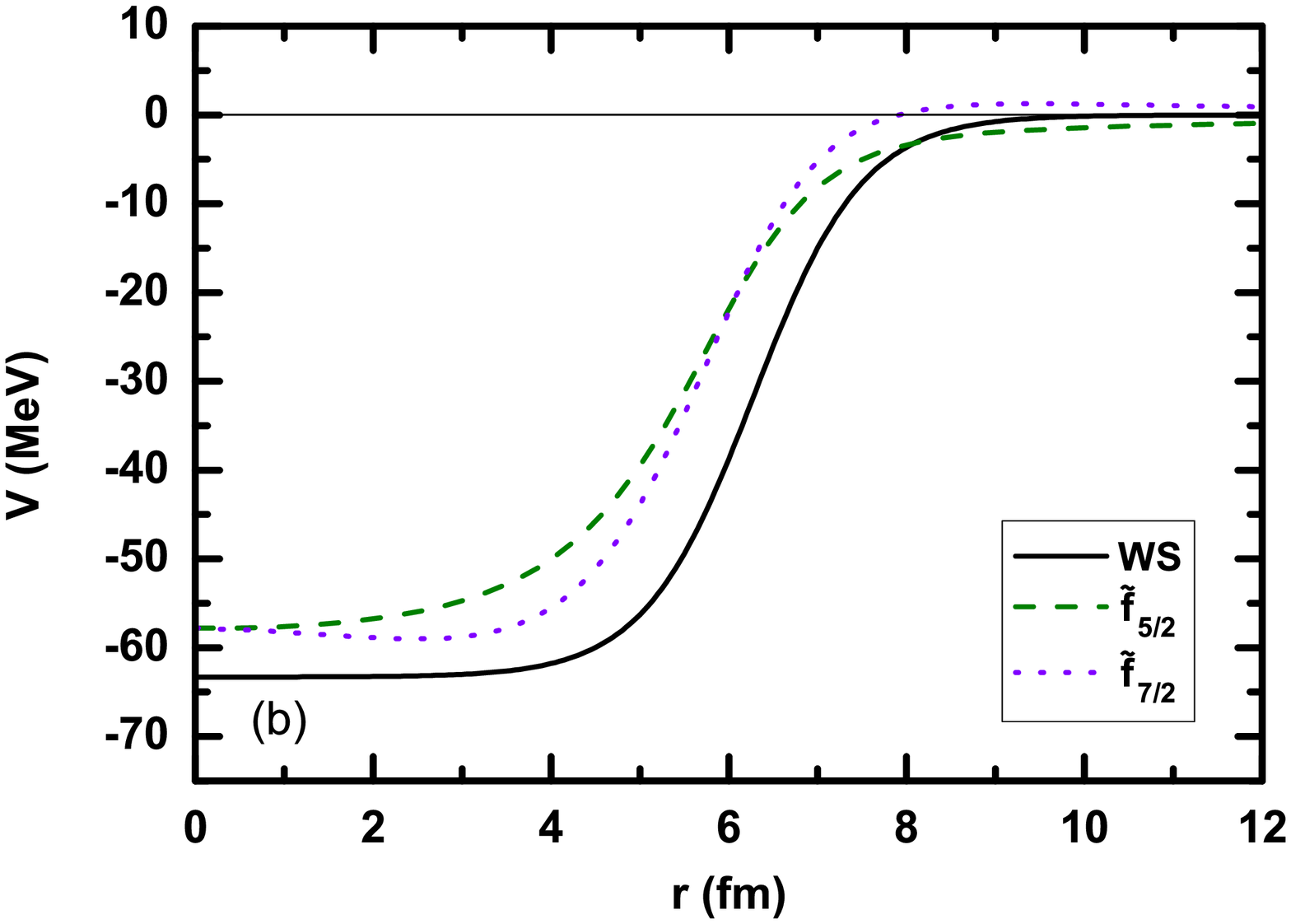}\\
\includegraphics[width=8cm]{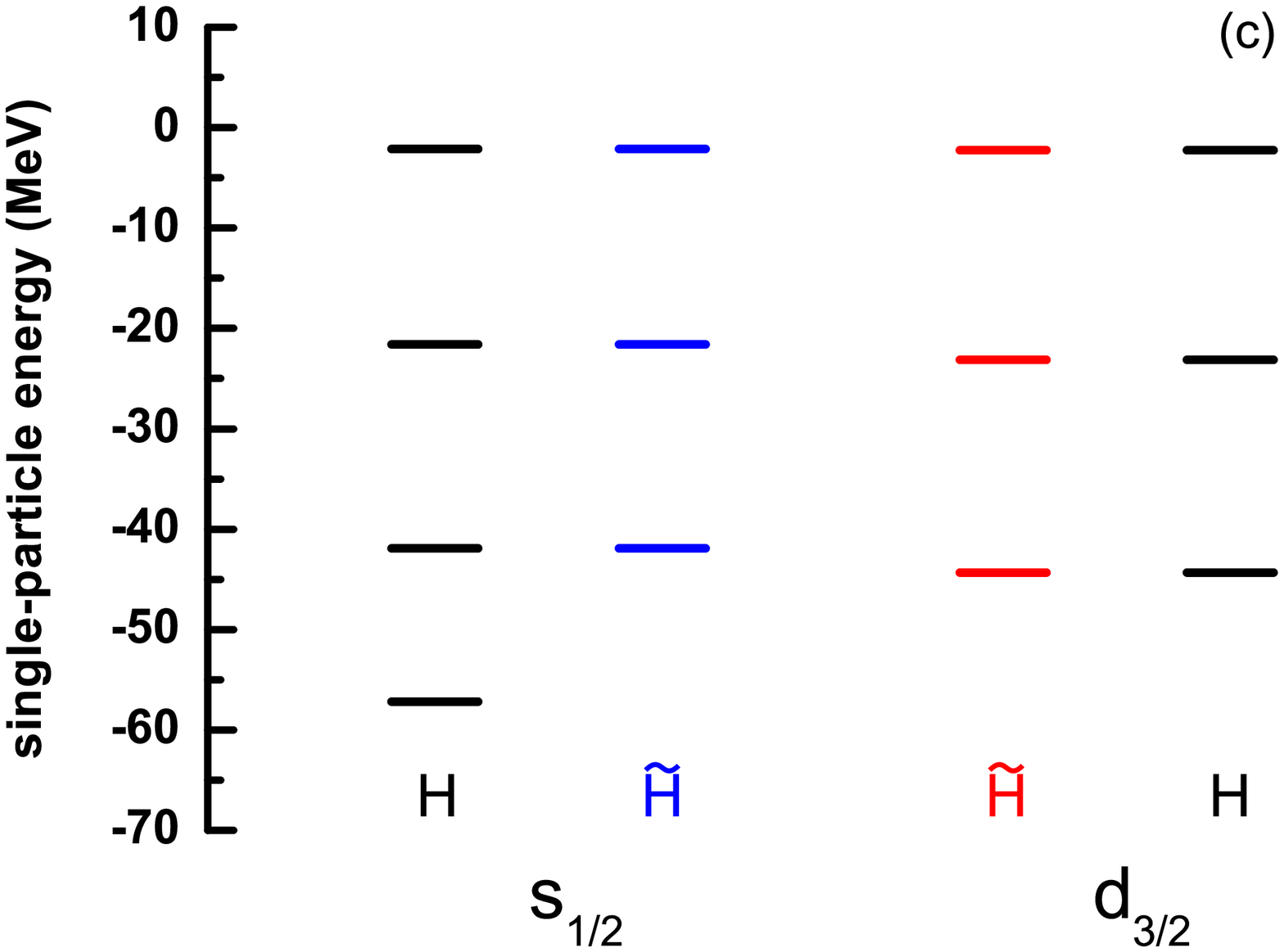}
\includegraphics[width=8cm]{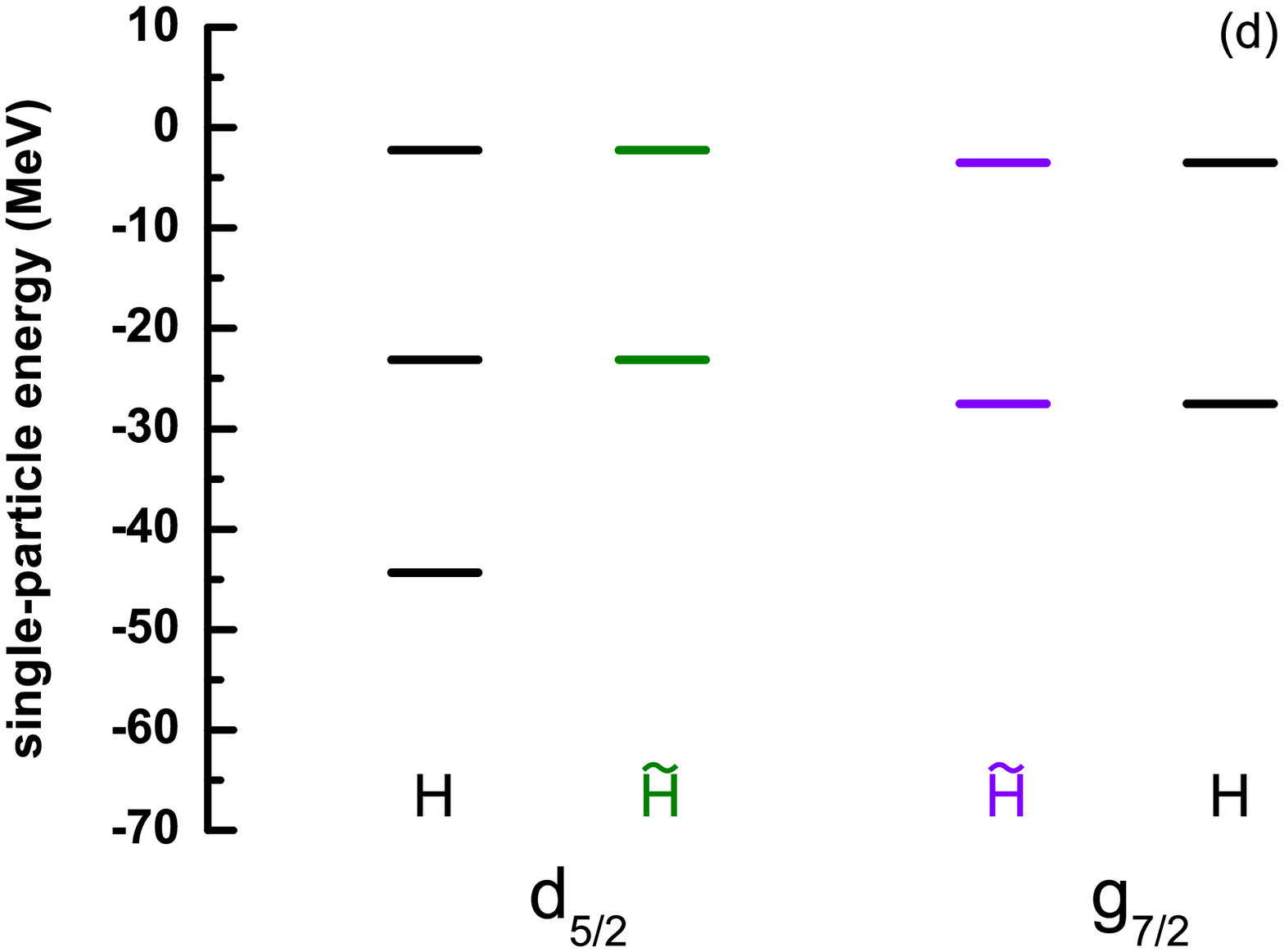}
\caption{(Color online) Upper panels: $\kappa$-dependent central potentials $\tilde{V}_\kappa(r)$ in $\tilde{H}$ as a function of $r$ for the $\tilde{p}$ and $\tilde{f}$ blocks, while the Woods-Saxon potential in $H$ is shown for comparison.
Lower panels: The corresponding single-particle energies obtained with $H$ and $\tilde{H}$.
    \label{Fig8}}
\end{figure*}

In the upper panels of Fig.~\ref{Fig8}, the central potentials $\tilde V_\kappa(r)$ in $\tilde{H}$ are shown by taking the $\tilde{p}$ and $\tilde{f}$ blocks as examples, while the Woods-Saxon potential $V(r)$ in $H$ is also shown for comparison.
For all blocks, the potentials $\tilde V_\kappa(r)$ approximately remain a Woods-Saxon shape, and they are shallower than the original potential $V(r)$.
Focusing on a pair of pseudospin partners, one sees that the potential $\tilde V_\kappa(r)$ with $\kappa<0$ is higher than that of its pseudospin partner with $\kappa>0$ at $r<6$~fm, since in this region the superpotential $q_\kappa(r)$ with $\kappa<0$ increases faster with $r$ as shown in Fig.~\ref{Fig7}.
In contrast, these two potentials reverse at $r>6$~fm, since in this region the slope of $q_\kappa(r)$ with $\kappa<0$ becomes negative whereas that with $\kappa<0$ remains positive.
By comparing the panels (a) and (b), it is seen that the amplitude of the difference between $\tilde V_\kappa(r)$ for a pair of pseudospin partners increases with the difference of their quantum numbers $|\kappa_a-\kappa_b|$.

After getting the central potentials $\tilde V_\kappa(r)$, we are ready to calculate the single-particle energies and wave functions of the SUSY partner Hamiltonians $\tilde{H}(\kappa)$.
In the lower panels of Fig.~\ref{Fig8}, the discrete single-particle energies obtained with $\tilde{H}$ are shown and compared with those obtained with $H$.
It is clearly shown that the eigenstates of Hamiltonians $H$ and $\tilde{H}$ are exactly one-to-one identical, except for the lowest eigenstates with $\kappa<0$ in $H$, which are the so-called intruder states.
In other words, the fact that the intruder states have no pseudospin partners can be interpreted as a natural result of the exact SUSY for $\kappa<0$ and broken SUSY for $\kappa>0$.

By holding the one-to-one mapping relation in the two sets of spectra, the origin of PSS, which is deeply hidden in the Hamiltonian $H$ within the normal scheme, can be now traced by employing its SUSY partner Hamiltonian $\tilde{H}$.

\begin{figure}
\includegraphics[width=8cm]{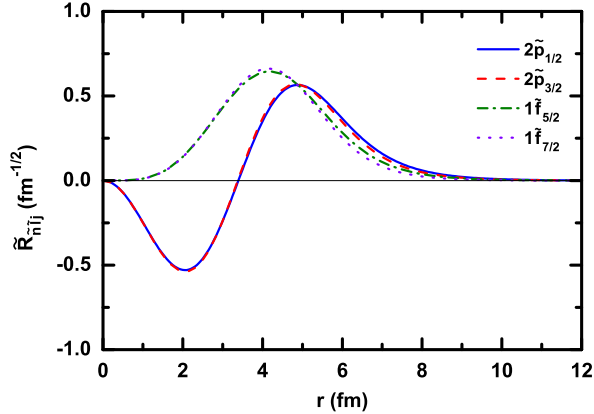}
\caption{(Color online) Single-particle wave functions $\tilde{R}_{\tilde{n}\tilde{l}j}(r)$ of $\tilde{H}$ for the $2\tilde{p}_{1/2}$ and $1\tilde{f}$ states.
    \label{Fig9}}
\end{figure}

In order to shed more light on this idea, we first show the single-particle radial wave functions $\tilde{R}_{\tilde{n}\tilde{l}j}(r)$ of $\tilde{H}$ for the $2\tilde{p}$ and $1\tilde{f}$ pseudospin doublets in Fig.~\ref{Fig9}.
In the SUSY quantum mechanics~\cite{Cooper1995}, the numbers of nodes in the radial wave functions $\tilde{R}_{\tilde{n}\tilde{l}j}(r)$ are one less than those in their counterparts $R_{nlj}(r)$ when the SUSY is exact, while the numbers of nodes in $\tilde{R}_{\tilde{n}\tilde{l}j}(r)$ are the same as those in their counterparts $R_{nlj}(r)$ when the SUSY is broken.
This indicates the well known node relation between the pseudospin doublets~\cite{Leviatan2001},
\begin{equation}\label{Eq:nodes}
    \tilde{n}=n-1\quad\mbox{for}\quad\kappa<0,\qquad
    \tilde{n}=n\quad\mbox{for}\quad\kappa>0,
\end{equation}
is nothing but an intrinsic property of the SUSY quantum mechanics.
This node relation can be also checked by comparing the wave functions shown in Fig.~\ref{Fig9} and Fig.~\ref{Fig4}.
In fact, not only are the numbers of nodes equal, but also the wave functions of pseudospin doublets are almost identical to each other.
Therefore, within this representation, the quasi-degeneracy of pseudospin doublets is closely related to the similarity of their wave functions, and vice versa.

\begin{table}
\caption{Contributions from kinetic term (kin.), pseudo-centrifugal barrier (PCB), and central potential (cen.) to the single-particle energies $E$ and the corresponding pseudospin-orbit splittings $\Delta E_{\rm PSO}$ for the pseudospin doublets $2\tilde{p}$ and $1\tilde{f}$.
All units are in MeV.}
\label{Tab2}
\begin{ruledtabular}
    \begin{tabular}{lrrrr}
        state & \multicolumn{1}{c}{$E_{\rm kin.}$} & \multicolumn{1}{c}{$E_{\rm PCB}$} & \multicolumn{1}{c}{$E_{\rm cen.}$} & \multicolumn{1}{c}{$E$} \\ \hline
        $2\tilde{p}_{1/2}$ & 16.602 &  6.723 & -44.916 & -21.591 \\
        $2\tilde{p}_{3/2}$ & 17.331 &  6.857 & -47.332 & -23.143 \\
      $\Delta E_{\rm PSO}$ & -0.729 & -0.134 &   2.415 &   1.552 \\ \hline
        $1\tilde{f}_{5/2}$ &  5.710 & 16.286 & -45.139 & -23.143 \\
        $1\tilde{f}_{7/2}$ &  6.293 & 16.591 & -50.392 & -27.508 \\
      $\Delta E_{\rm PSO}$ & -0.584 & -0.305 &   5.253 &   4.365 \\
    \end{tabular}
\end{ruledtabular}
\end{table}

The same strategy is then adopted to investigate the PSO splittings $\Delta E_{\rm PSO}$ by decomposing the contributions from each term as done in Table~\ref{Tab1}, but now within the representation of $\tilde{H}$ shown in Eq.~(\ref{Eq:Htil}) instead.
The corresponding operators include the kinetic term $-d^2/(2Mdr^2)$, the pseudo-centrifugal barrier (PCB) $\kappa(\kappa-1)/(2Mr^2)$, and the central potential $\tilde{V}_\kappa(r)$.
The corresponding results for the pseudospin doublets $2\tilde{p}$ and $1\tilde{f}$ are listed in Table~\ref{Tab2}.
It can be seen that for each pair of pseudospin doublets the energy contributions from the PSS conserving terms, i.e., the kinetic term and PCB, are very similar.
The PSO splittings $\Delta E_{\rm PSO}$ are mainly contributed by the difference in the central potentials $\Delta E_{\rm cen.}$, which is due to the slight $\kappa$-dependence of $\tilde{V}_\kappa(r)$ as shown in Fig.~\ref{Fig8}.
In other words, the sophisticated cancellations among different terms in $H$ can be clearly understood
by using a proper decomposition with the help of the SUSY quantum mechanics.

\begin{figure}
\includegraphics[width=8cm]{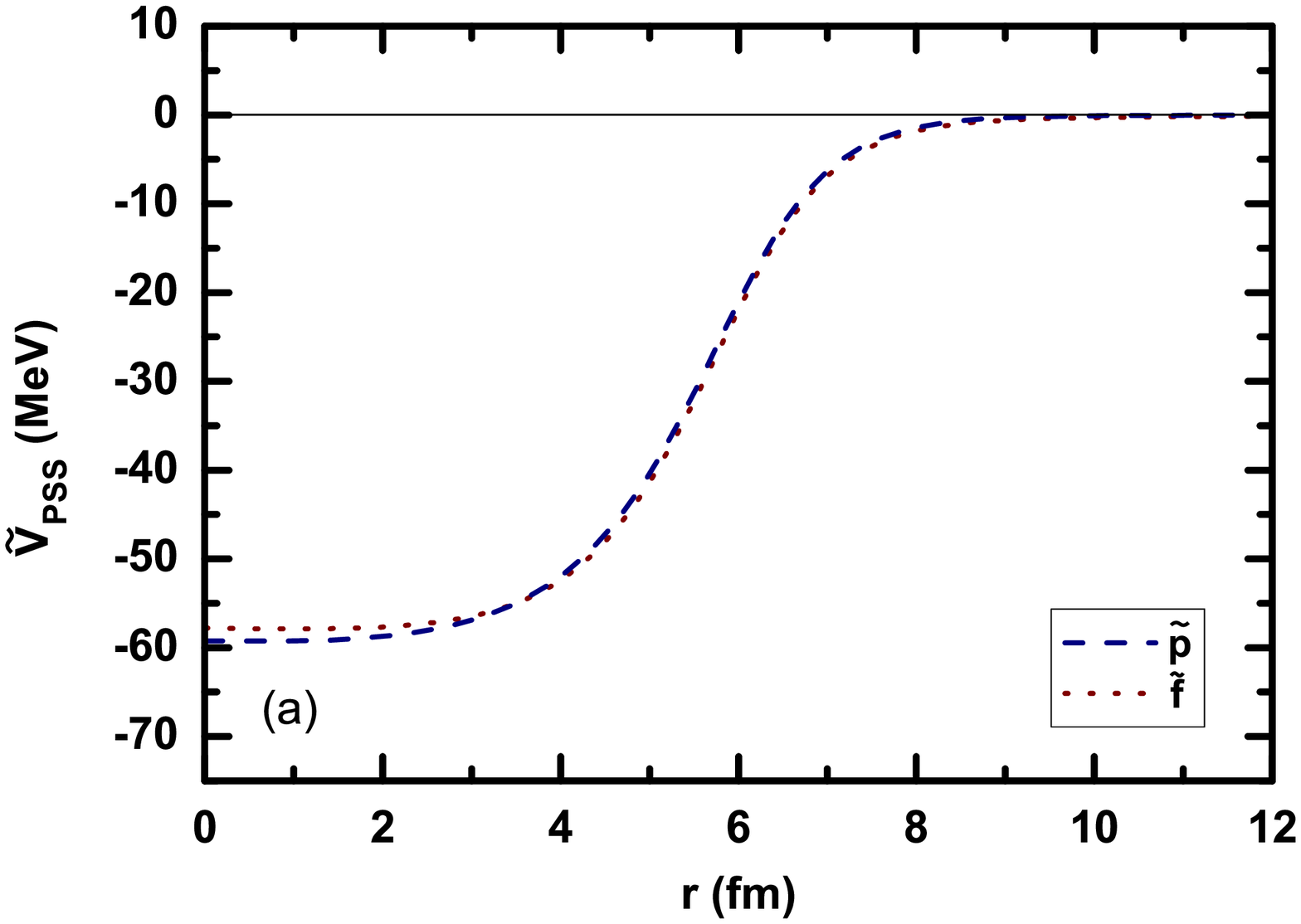}\\
\includegraphics[width=8cm]{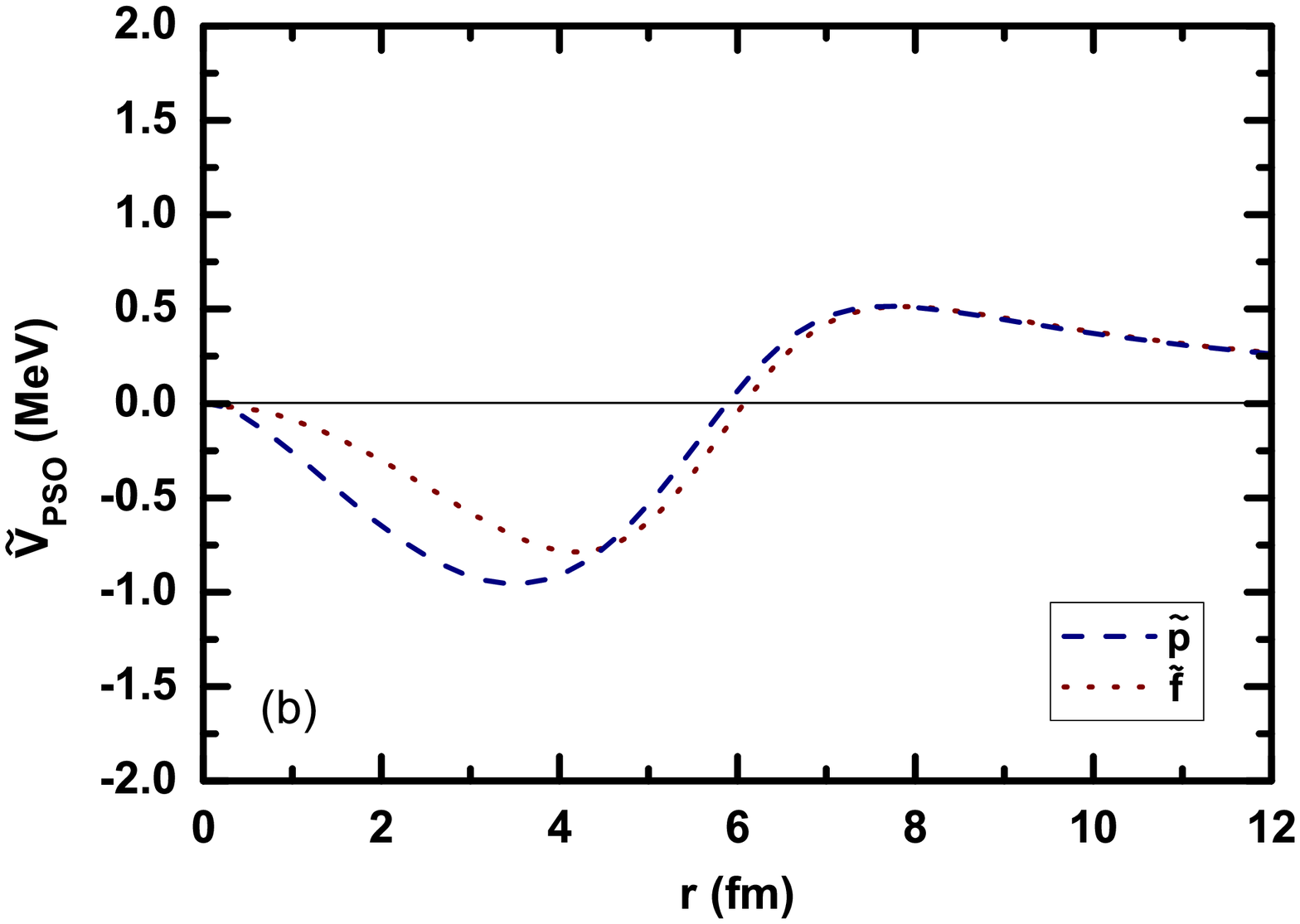}
\caption{(Color online) Pseudospin symmetry conserving potentials $\tilde{V}_{\rm PSS}(r)$ (a) and breaking potentials $\tilde{V}_{\rm PSO}(r)$ (b) for the $\tilde{p}$ and $\tilde{f}$ blocks.
    \label{Fig10}}
\end{figure}

In order to perform the quantitative perturbation calculations, the Hamiltonian $\tilde{H}$ is split as
\begin{equation}
  \tilde{H} = \tilde{H}^{\rm PSS}_0 + \tilde{W}^{\rm PSS},
\end{equation}
where $\tilde{H}^{\rm PSS}_0$ and $\tilde{W}^{\rm PSS}$ are the corresponding PSS conserving and breaking terms, respectively.
By requiring that $\tilde{W}^{\rm PSS}$ should be proportional to $\kappa$, which is similar to the case of the spin-orbit term in the normal scheme, one has
\begin{subequations}
\begin{eqnarray}
  \tilde{V}_{\kappa_a}(r) &=& \tilde{V}_{\rm PSS}(r) + \kappa_a \tilde{V}_{\rm PSO}(r),\\
  \tilde{V}_{\kappa_b}(r) &=& \tilde{V}_{\rm PSS}(r) + \kappa_b \tilde{V}_{\rm PSO}(r),
\end{eqnarray}
\end{subequations}
for a pair of pseudospin doublets with $\kappa_a+\kappa_b=1$.
In such a way, the PSS conserving potentials $\tilde{V}_{\rm PSS}(r)$ and breaking potentials $\tilde{V}_{\rm PSO}(r)$ can be uniquely determined as
\begin{subequations}
\begin{eqnarray}
  \tilde{V}_{\rm PSS}(r) &=& \frac{\kappa_b\tilde{V}_{\kappa_a}(r) - \kappa_a\tilde{V}_{\kappa_b}(r)}{\kappa_a-\kappa_b},\\
  \tilde{V}_{\rm PSO}(r) &=& \frac{\tilde{V}_{\kappa_a}(r) - \tilde{V}_{\kappa_b}(r)}{\kappa_a-\kappa_b}
    =\frac{1}{M}\frac{q'_{\kappa_a}(r) - q'_{\kappa_b}(r)}{\kappa_a-\kappa_b}.
\end{eqnarray}
\end{subequations}
The corresponding $\tilde{H}^{\rm PSS}_0$ and $\tilde{W}^{\rm PSS}$ read
\begin{subequations}
\begin{eqnarray}
    \tilde{H}^{\rm PSS}_0&=&\ff{2M}\ls-\frac{d^2}{dr^2}+\frac{\kappa(\kappa-1)}{r^2}\rs + \tilde{V}_{\rm PSS}(r),\\
    \tilde{W}^{\rm PSS}&=&\kappa \tilde{V}_{\rm PSO}(r).
\end{eqnarray}
\end{subequations}

By taking the $\tilde{p}$ and $\tilde{f}$ blocks as examples, the PSS conserving potentials $\tilde{V}_{\rm PSS}(r)$ and breaking potentials $\tilde{V}_{\rm PSO}(r)$ are illustrated in Fig.~\ref{Fig10}.
On one hand, it can be seen that the PSS conserving potentials $\tilde{V}_{\rm PSS}(r)$ remain an approximate Woods-Saxon shape, and they are $\kappa$-dependent to a small extent.
On the other hand, the PSS breaking potentials $\tilde{V}_{\rm PSO}(r)$ show several special features.
First of all, these PSS breaking potentials are regular functions of $r$, in particular, they vanish at $r\rightarrow\infty$.
This was also one of the main goals of the investigations in Ref.~\cite{Typel2008}, but here we not only achieve the goal, but also keep every operator Hermitian.
Second, it can be seen that the amplitudes of $\tilde{V}_{\rm PSO}$ are around 1~MeV, which directly lead to the reduced PSO splittings $\Delta E_{\rm PSO}\lesssim1$~MeV as shown in Fig.~\ref{Fig3}.
More importantly, different from the usual SO potentials with a surface-peak shape, the PSO potentials $\tilde{V}_{\rm PSO}(r)$ are negative at small radius but positive at large radius with a node at the surface region.
This property can be traced back to the differential equation (\ref{Eq:HV}) of the reduced superpotentials $q_\kappa(r)$.
By analyzing the corresponding asymptotic behaviors of $q_\kappa(r)$ at $r\rightarrow0$ and $r\rightarrow\infty$, one can conclude that such particular shape of $\tilde{V}_{\rm PSO}(r)$ holds as long as the central potentials $V(r)$ in the Schr\"odinger equations are of a Woods-Saxon-like shape.

The particular shape of the PSO potentials $\tilde{V}_{\rm PSO}(r)$ can explain well the variations of the PSO splitting with the single-particle energy.
First of all, it has been emphasized above that $\tilde{V}_{\rm PSO}(r)$ do not depend on the radial quantum number $n$.
Meanwhile, the single-particle wave functions $\tilde{R}(r)$ extend to larger distance with higher energies.
Thus, the matrix element $\langle\tilde{R}|\tilde{V}_{\rm PSO}|\tilde{R}\rangle$ is negative when the wave function is centralized in the inner part.
As the wave function becomes more extended, the positive part of $\tilde{V}_{\rm PSO}(r)$ compensates for the negative value of the matrix element.
In such a way, the PSO splittings $\Delta E_{\rm PSO}$ decrease while the radial quantum numbers $\tilde{n}$ increase.
The splittings can even reverse for the resonance states, where the outer part of the PSO potentials plays the major role.

\begin{figure}
\includegraphics[width=8cm]{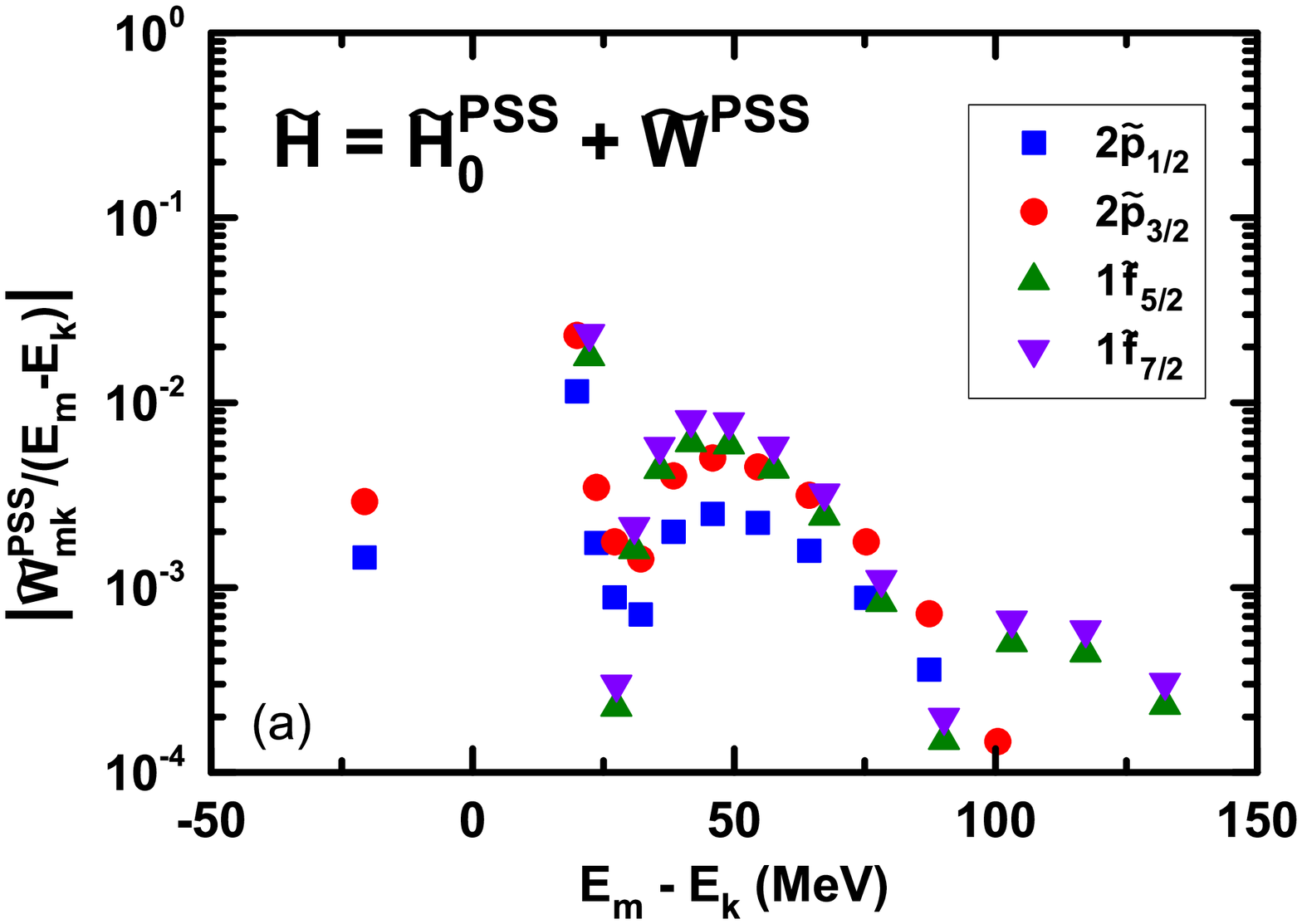}\\
\includegraphics[width=7.55cm]{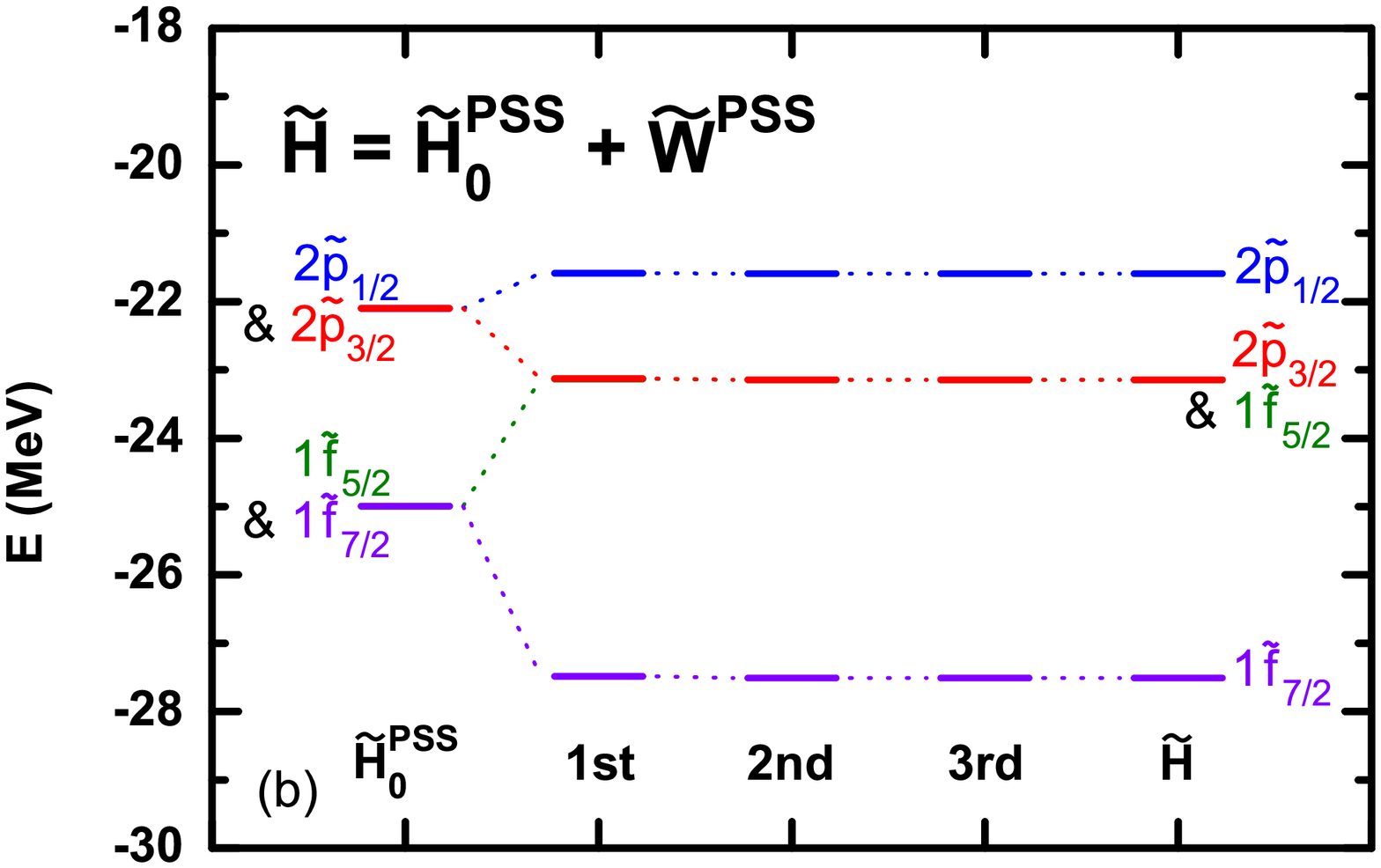}
\caption{(Color online) Same as Fig.~\ref{Fig6}, but for the case of $\tilde{H}=\tilde{H}^{\rm PSS}_0+\tilde{W}^{\rm PSS}$.
    \label{Fig11}}
\end{figure}

Finally, we perform the perturbation calculations based on the pseudospin symmetric Hamiltonian $\tilde{H}^{\rm PSS}_0$ with the perturbation $\tilde{W}^{\rm PSS}$.
In Fig.~\ref{Fig11}, the values of $\left|\tilde{W}^{\rm PSS}_{mk}/(E_m-E_k)\right|$ for the pseudospin doublets $k = 2\tilde{p}$ and $1\tilde{f}$ are shown as functions of the energy differences $E_m-E_k$ in the upper panel, while the single-particle energies obtained at the PSS limit $\tilde{H}^{\rm PSS}_0$, and their counterparts obtained by the first-, second-, and third-order perturbation calculations, as well as those obtained with $\tilde{H}$ are shown from left to right in the lower panel.
It can be seen that the pseudospin doublets are exactly degenerate at the PSS limit $\tilde{H}^{\rm PSS}_0$.
It can also be examined that the radial wave functions for each pair of pseudospin doublets are identical.
For the present decomposition, the largest perturbation correction is less than 0.03,
which is almost one order of magnitude smaller than that with the decomposition $H=H^{\rm HO}_0+W^{\rm HO}$.
This indicates that the criterion in Eq.~(\ref{Eq:criterion}) is satisfied quite well.
As shown in the lower panel, not only the PSO splittings but also the energy degeneracy of the spin doublets are excellently reproduced by the first-order perturbation calculations.

In such an explicit and quantitative way, the PSO splittings $\Delta E_{\rm PSO}$ can be directly understood by the PSS breaking term $\tilde{W}^{\rm PSS}$ within the representation of the SUSY partner Hamiltonian $\tilde{H}$. Furthermore, this symmetry breaking term can be treated as a very small perturbation on the exact PSS limit $\tilde{H}^{\rm PSS}_0$. Therefore, the PSS in the present potential without spin-orbit term is of pertubative nature..

\section{Summary and Perspectives}\label{Sect:IV}

The supersymmetric (SUSY) quantum mechanics is used to investigate the origin of the pseudospin symmetry (PSS) and its breaking in the single-particle spectra in nuclei.
In the formalism, it can be seen that, while the spin symmetry conserving term $\kappa(\kappa+1)$ appears in the single-particle Hamiltonian $H$, the PSS conserving term $\kappa(\kappa-1)$ appears naturally in its SUSY partner Hamiltonian $\tilde{H}$.
In addition, the fact that all states with $\tilde{l}>0$, except for intruder states, have their own pseudospin partners can be interpreted by employing both exact and broken SUSY in a unified way.

In the present study, we focus on a Schr\"odinger equation with a Woods-Saxon central potential.
This corresponds to the lowest-order approximation for transforming a Dirac equation into a diagonal form by using the similarity renormalization group (SRG) technique.

Within the single-particle Hamiltonian $H$ representation, the wave functions of pseudospin doublets are very different from each other, and the small pseudospin-orbit (PSO) splittings $\Delta E_{\rm PSO}$ are due to the sophisticated cancellations among all channels.
However, this does not necessarily mean the dynamical or even the nonperturbative nature of PSS.
By taking the Hamiltonian with a harmonic oscillator (HO) potential $H^{\rm HO}_0$ as the exact PSS limit, it is found that the largest perturbation correction due to the symmetry breaking term $W^{\rm HO}$ is roughly 0.13, and $\Delta E_{\rm PSO}$ can be well reproduced by the third-order perturbation calculations.

The origin of the PSS and its breaking mechanism can be interpreted explicitly within the representation of the SUSY partner Hamiltonian $\tilde{H}$.
In order to obtain $\tilde{H}$, one first solves the first-order differential equation for the reduced superpotentials $q_\kappa(r)$, and then one obtains the $\kappa$-dependent central potentials $\tilde{V}_\kappa(r)$ in $\tilde{H}$.
It is important that the central potentials are regular and converge at both $r\rightarrow0$ and $r\rightarrow\infty$.
Consequently, the eigenstates of Hamiltonians $H$ and $\tilde{H}$ are exactly one-to-one identical, except for the additional eigenstates in $H$ when the SUSY is exact, which correspond to the intruder states without pseudospin partners.

By holding this one-to-one mapping relation, the origin of PSS deeply hidden in $H$ can be traced by employing its SUSY partner Hamiltonian $\tilde{H}$.
Within this SUSY partner scheme, the wave functions of the pseudospin doublets are almost identical, and the well known node relation between the pseudospin doublets, i.e., $\tilde{n}=n-1$ for $\kappa<0$ and $\tilde{n}=n$ for $\kappa>0$, is one of the intrinsic properties of the SUSY quantum mechanics.
It is also found that the PSO splittings $\Delta E_{\rm PSO}$ are mainly contributed by the PSS breaking terms.
This indicates the sophisticated cancellations among different terms in $H$ can be understood in a clear scheme by using a proper decomposition.

The Hamiltonian $\tilde{H}$ is then split into the PSS conserving term $\tilde{H}^{\rm PSS}_0$ and the PSS breaking term $\tilde{W}^{\rm PSS}$.
While the PSS conserving potentials $\tilde{V}_{\rm PSS}(r)$ remain an approximate Woods-Saxon shape with a slight $\kappa$-dependence, the PSS breaking potentials $\tilde{V}_{\rm PSO}(r)$ show several special features.
1) The potentials $\tilde{V}_{\rm PSO}(r)$ are regular functions of $r$, and they vanish at both $r\rightarrow0$ and $r\rightarrow\infty$.
2) The amplitudes of $\tilde{V}_{\rm PSO}$ are around 1~MeV, which directly lead to the reduced PSO splittings $\Delta E_{\rm PSO}\lesssim1$~MeV.
3) The potentials $\tilde{V}_{\rm PSO}(r)$ show a particular shape of being negative at small radius but positive at large radius, with a node at the surface region.
The general pattern that the PSO splittings become smaller with increasing single-particle energies, and even reverse for resonance states, can be understood straightforwardly by such particular symmetry-breaking potentials.

Finally, the perturbation calculations are performed based on the PSS conserving Hamiltonian $\tilde{H}^{\rm PSS}_0$ with the perturbation $\tilde{W}^{\rm PSS}$.
It is found that in such decomposition the largest perturbation correction due to the symmetry breaking term $\tilde{W}^{\rm PSS}$ is less than 0.03, and not only the PSO splittings but also the energy degeneracy of the spin doublets are excellently reproduced by the first-order perturbation calculations.
In such a way, the origin of the PSS can be recognized, and its breaking is due to a very small perturbation $\tilde{W}^{\rm PSS}$ on the exact PSS limit $\tilde{H}^{\rm PSS}_0$.

In short, the justification for using SUSY quantum mechanics is that the partner Hamiltonian shares its eigenvalues with the original one, while the PSS conserving term proportional to $\kappa(\kappa-1)$ can be naturally identified. Furthermore, the PSS breaking term is responsible for the observed PSS splitting. The amplitudes of $\tilde V_{\rm PSO}$ quantitatively determine the amplitudes of the reduced PSO splittings $\Delta E_{\rm PSO}$. The particular shape of $\tilde V_{\rm PSO}(r)$ can also explain the decrease of the PSO splitting with increasing single-particle energies.

The present investigation employs a Schr\"odinger equation for illustrating the key ideas on applying the SUSY quantum mechanics to the PSS in nuclei. Since the spin-orbit term, which appears as a second-order correction in $1/M$, is crucial for the nuclear shell structure, it is important to investigate its effects on the properties of the SUSY quantum mechanics and PSS in a quantitative way. In order to completely answer the question of why the PSS is conserved better than the SS in realistic nuclei, the intrinsic relation between the spin-orbit potential and the central potential or the effective mass must be taken into account.
In this sense, the PSS must be regarded as the relativistic symmetry, and it should be recognized in the Dirac equation, or equivalently the Schr\"odinger-like equation obtained by transforming the Dirac equation into a diagonal form by using the SRG technique.

\section*{ACKNOWLEDGMENTS}
This work was partly supported by the Major State 973 Program
2013CB834400, the NSFC under Grants No. 10975008, No. 11105005, No. 11105006, and No. 11175002,
China Postdoctoral Science Foundation Grants No. 20100480149 and No. 201104031, and the Research Fund for the Doctoral Program of Higher Education under Grant No. 20110001110087.


\end{CJK*}
\end{document}